\documentclass[journal,draftclsnofoot,onecolumn,12pt]{IEEEtran}

\ifCLASSINFOpdf
\else
\fi

\usepackage{cite}
\usepackage{graphicx}
\usepackage{multirow}
\usepackage{amsfonts}
\usepackage{amssymb}
\usepackage{booktabs}
\usepackage{siunitx}
\usepackage[table]{xcolor}
\usepackage{amsmath}
\usepackage{array}

\usepackage{algorithm}
\usepackage{algpseudocode}
\usepackage{array}
\usepackage{dsfont}
\usepackage{makecell}

\usepackage[font=small,skip=10pt]{caption}
\usepackage{siunitx}
\usepackage{booktabs, makecell}
\usepackage[labelformat=brace]{subcaption}
\DeclareCaptionFormat{myformat}{\hspace*{1.2cm}#1#2#3}
\usepackage{color}
\usepackage{float}
\usepackage{tikz}
\usepackage[utf8]{inputenc}
\usepackage{pgfplots} 
\usepackage{pgfgantt}
\usepackage{pdflscape}
\pgfplotsset{compat=newest} 
\pgfplotsset{plot coordinates/math parser=false}
\newlength\fwidth

\def\sinr{\mbox{\small$\mathsf{SINR}$}}

\begin{document}

\title{Joint Uplink-Downlink Capacity and Coverage Optimization via Site-Specific Learning of Antenna Settings}

 \author{Ezgi~Tekgul, Thomas~Novlan, Salam~Akoum, Jeffrey G.~Andrews
  \thanks{Ezgi Tekgul and Jeffrey G. Andrews are with 6G@UT in the Wireless Networking and Communications Group and the Dept. of Electrical and Computer Engineering at the University of Texas at Austin, Austin,
  TX, USA (e-mail: ezgitekgul@utexas.edu, jandrews@ece.utexas.edu), Thomas Novlan and Salam Akoum are with AT{\&}T Labs, Austin, TX, USA (e-mail: tn911r@att.com, sa469y@att.com). Preliminary version appeared in IEEE ITW \cite{tekgul2021sample}.  Date last modified: \today.}
 }

\maketitle

\begin{abstract}
We propose a novel framework for optimizing antenna parameter settings in a heterogeneous cellular network. We formulate an optimization problem for both coverage and capacity -- in both the downlink (DL) and uplink (UL) -- which configures the tilt angle, vertical half-power beamwidth (HPBW), and horizontal HPBW of each cell’s antenna array across the network. The novel data-driven framework proposed for this non-convex problem, inspired by Bayesian optimization (BO) and differential evolution algorithms, is sample-efficient and converges quickly, while being scalable to large networks. By jointly optimizing DL and UL performance, we take into account the different signal power and interference characteristics of these two links, allowing a graceful trade-off between coverage and capacity in each one.  Our experiments on a state-of-the-art 5G NR cellular system-level simulator developed by AT{\&}T Labs show that the proposed algorithm consistently and significantly outperforms the 3GPP default settings, random search, and conventional BO. In one realistic setting, and compared to conventional BO, our approach increases the average sum-log-rate by over $60\%$ while decreasing the outage probability by over $80\%$.  Compared to the 3GPP default settings, the gains from our approach are considerably larger.  The results also indicate that the practically important combination of DL throughput and UL coverage can be greatly improved by joint UL-DL optimization.  
\end{abstract}


%
\IEEEpeerreviewmaketitle

\section{Introduction}
\subsection{Motivation}
Cellular system capacity and coverage depend on the base station (BS) antenna settings, i.e. the shape and direction of the dominant beams.  Well-tuned antenna parameters -- namely tilt angle, vertical half-power beamwidth (HPBW), and horizontal HPBW -- increase the received signal strength over key areas of the cell and minimize interference to neighboring cells.  The optimization of antenna parameters across a network is nontrivial because the settings across each cell are coupled by the interference, rendering the multicell optimization problem non-convex and NP-hard \cite{luo2008dynamic}.   Another challenge stems from the conflicting nature of the two key objectives, maximizing both coverage probability -- which in practice means directing energy towards the cell edges at the expense of other-cell interference -- and the sum or more often sum-log capacity, which tends to favor cell interior users with high SINR.   Finally, the eventual configurations should balance between uplink and downlink performance, which has rarely been considered.  Standard optimization approaches tend to be prohibitively complex and ineffective for even modest network sizes.  

In the Third Generation Partnership Project (3GPP), global optimization methods based on stochastic system simulation are utilized to optimize parameter settings. Since the network models are usually small homogeneous hexagonal layouts, exhaustive search techniques can be used, resulting in typical fixed values that are the same for all cells, e.g., $12^{\circ}$ downtilt angles. In a real network, the antenna parameters can be designed using site-specific radio frequency planning tools, and any updates rely on trial-and-error methods based on field measurements over a long time period. These methods are neither scalable nor near-optimal, and hence there is a need for practical and automated optimization approaches, that are well-supported by theory and utilizes recent advances in data-driven design.   This paper proposes such a framework.
\vspace{-0.2cm}
\subsection{Prior Work}
The antenna parameter-based coverage and capacity optimization (CCO) literature has focused on downlink (DL) optimization: uplink (UL) optimization has received very little attention. However, the optimal DL antenna parameters are generally suboptimal for the UL due to the major signal power and interference asymmetries between two links. The DL interference is dictated by the fixed locations of BSs which transmit nearly continuously using high gain antennas, while in the uplink, mobile user equipments (UEs) which each transmits sporadically and (usually) omnidirectionally generate the interference. Furthermore, large transmit power disparities exist between different types of BSs in a heterogeneous network (HetNet), resulting in coverage area differences, whereas UEs transmit at relatively low power, with location-dependent power control, amplifying the difference between the interference characteristics of the uplink and the downlink \cite{novlan2013analytical, singh2015joint}. Given the importance of uplink coverage even for DL-centric data traffic -- DL transmissions are not possible without reliable UL control channels -- it is clear that UL coverage in particular should be considered when optimizing the antenna settings.   Despite this, to the best of our knowledge, \cite{berger2015joint} is the only work in the literature that considers joint UL/DL antenna parameter-based CCO.   They consider only downtilt under sparse system knowledge, however, and their descent-like search approach can only optimize a single network parameter.

Various studies parameterized by the key antenna parameters including transmit power \cite{fan2014self, engels2013autonomous}, downtilt \cite{razavi2010self, ul2012cooperative, engels2013autonomous, berger2014online, buenestado2016self}, azimuth \cite{awada2011optimizing, kasem2013antenna}, and HPBW \cite{yilmaz2009analysis, kasem2013antenna} have been conducted on downlink-only CCO.   They have mainly focused on intelligent networks after the introduction of self-organizing network functionalities in 4G networks \cite{aliu2012survey}. Among these works, \cite{awada2011optimizing, engels2013autonomous, kasem2013antenna, berger2014online, buenestado2016self} use traditional optimization and rule-based approaches, which are poor at adapting to different environments and thus require manual intervention. On the other hand, \cite{razavi2010self, ul2012cooperative, fan2014self} combine rule-based fuzzy systems with reinforcement learning (RL) for a more adaptive implementation. However, this fuzzy RL method struggles to handle continuous or high-dimensional network configurations and leads to a complicated process for determining the reward signal for each state-action pair, due to the activation of multiple fuzzy membership functions at the fuzzification stage \cite{razavi2010self}. 

More recent methods that have been used for antenna tuning are RL \cite{dandanov2017dynamic, balevi2019online, dreifuerst2021optimizing, vannella2021remote, bouton2021coordinated} and Bayesian optimization (BO) \cite{dreifuerst2021optimizing}. RL superficially may seem to be a suitable approach for CCO with its ability to adapt to changing environmental dynamics. However, RL methods need a very large amount of data to achieve high accuracy and tend to have slow convergence, resulting in extensive computations and long-lasting simulations \cite{benzaid2020ai}, a fact we have encountered in our own studies over the last few years.   RL also lacks safe exploration as random (e.g., epsilon-greedy) exploration can result in undesirable antenna parameters being tested which significantly degrade the system performance.  Furthermore, it is not advisable or even possible to make large sudden changes in the parameters, especially the tilt angle. Alternatively, if just small incremental changes are allowed at each iteration, then it further slows convergence, which is then often to a local maximum far below the global maxima.  

BO is a more promising approach for the CCO problem, and it can speed up convergence and provide safe exploration \cite{maggi2021bayesian}. In \cite{dreifuerst2021optimizing}, the deep deterministic policy gradient algorithm (DDPG), an RL method, was compared with BO for optimizing network coverage and capacity, and it was shown that BO improves sample efficiency by over two orders of magnitude relative to DDPG.  However, neither of these two approaches scales well with network size.  BO suffers from cubic computational complexity, which limits its applications to low-dimensional problems. It decides on the next sampling point with a so-called acquisition function, and this requires solving a non-convex problem with increasing computational cost as the training data size increases at each iteration \cite{maggi2021bayesian}. This also makes choosing a proper acquisition function a challenging task in the implementation of BO as it has a major impact on its performance.
\vspace{-0.2cm}
\subsection{Contributions}
We investigate the joint uplink and downlink antenna parameter optimization problem and propose a data-driven method for fast site-specific optimization of capacity and coverage. The proposed novel algorithm leverages several key aspects of both Bayesian optimization (BO) and differential evolution (DE) and has several practically desirable properties which we now enumerate.

\textbf{Sample-and time-efficient novel algorithm.} Our proposed approach is inspired by BO's probabilistic model-building, which we combine with an  evolutionary algorithm to quickly search and prune the space of candidate solutions.   This is a sample-and time-efficient framework, improving sample efficiency by over two orders of magnitude compared to DDPG.  Furthermore, our algorithm is demonstrated to have linear time complexity as opposed to cubic in BO. 

\textbf{Scalable to large network sizes.} Our approach gracefully scales to large network sizes and accommodates very large numbers of users while maintaining low complexity and high accuracy. To achieve this, we utilize the concept of a local \textit{neighborhood} for each user, which includes cells with large measured reference signal received power (RSRP). The SINR of each UE is then regressed on the parameters of the selected cells before calculating a cumulative metric for optimization, resulting in higher prediction accuracy. This also preserves the time efficiency of the algorithm by enabling the modeling part of the algorithm to be performed independently and thus in parallel for each user.

\textbf{Uplink and downlink joint optimization.} We jointly optimize the UL and DL directions of a multi-cell system, providing a framework that can trade off their importance as well as between coverage and capacity in each direction. We adjust the trade-off coefficient to focus on data rates in the downlink and coverage in the uplink. We then show that joint uplink and downlink optimization greatly improves the throughput and coverage performance of uplink (downlink) compared to the downlink-only (uplink-only) optimization with only a small loss from the uplink-only (downlink-only) optimization.

\textbf{Validation on a high-fidelity system-level simulator.}
We experimentally evaluate the performance of our proposed algorithm on a state-of-the-art wireless simulator developed by AT{\&}T Labs by comparing it with three baselines: (i) the default settings in 3GPP, (ii) random search, and (iii) conventional BO. This powerful system-level simulator closely mimics a real-world network, and we use a layout based on the real locations of AT{\&}T BSs to compare the algorithms in terms of sample and time efficiency, and three other performance metrics: (i) average sum-log-rate of the UEs, (ii) outage probability (i.e., the fraction of UEs with SINR value below a predefined threshold), and (iii) the  SINR  distribution  for  all  UEs.
\subsection{Notation and Organization}
We use bold lower-case and upper-case symbols to denote vectors and matrices, respectively. $\boldsymbol{X}^{T}$ denotes the transpose of matrix $\boldsymbol{X}$ and $\boldsymbol{x}_{i}$ denotes the $i$-th instance of vector $\boldsymbol{x}$. The notation $\mathbb{R}_{+}^{1 \times d}$ is used to represent $d$-dimensional vector of positive real numbers and $\mathbb{R}^{1 \times d}$ represents $d$-dimensional vector of real numbers. $\boldsymbol{1}_{D}$ denotes $D$-dimensional vector of all ones and $\mathds{1}\{\mathcal{A}\}$ denotes
the indicator function over set $\mathcal{A}$.

The rest of the paper is organized as follows. A system model is described in Section \ref{sec:System_Model}, and the optimization problem is formulated in Section \ref{sec:Problem_Formulation}. The proposed framework is presented with the evaluation metrics in Section \ref{sec:Framework}. The simulation details are given in Section \ref{sec:Simulation}, followed by the optimization performance results and discussions for the single and joint direction optimization in Section \ref{sec:Results_Discussion}. Conclusions and future directions are provided in Section \ref{sec:Conc_FW}.
%
%
\section{System Model and Problem Formulation} 
\subsection{System Model}
\label{sec:System_Model}
We consider a two-tier HetNet cellular network model consisting of a total of $M$ cells.  There are tower-mounted ``macro" BSs with three sector antennas, which can be deployed on a conventional hexagonal grid or using actual deployment locations: we consider both cases, including a current AT\&T network deployment.  The second tier are small cell (i.e. ``pico'') BSs which are randomly distributed within these layouts, adhering to a minimum distance (i.e., $10$m) between them and the macro BSs. We consider $N \gg M$ uniformly distributed UEs that are associated with a single BS based on the maximum downlink RSRP, which in effect corresponds to the BS with the minimum path loss (small cell biasing is not considered in this work).   Most of the system parameters including the transmit power, antenna gains, and channel and path loss models are derived from the 3GPP urban macro-cellular scenario described in \cite{3gpp.38.901} for a bandwidth of $10$ MHz at a carrier frequency of $2$ GHz.  The uplink transmit power obeys the 3GPP fractional power control scheme and is proportional to the bandwidth and the path loss.  This uplink transmit power is calculated as
\begin{equation*}
    P_{t} = \min(P_{\max}, 10\log(\mathcal{M}) + P_{0} + \varphi L)
\end{equation*} 
where $L$ is the downlink path loss, $P_{0}$ is the target power 
at the BS, $P_{\max}$ is the maximum uplink transmit power, and $\varphi$ is the fractional path loss compensation factor whose typical values are between $0.5$ and $0.9$: we use $\varphi = 0.9$. All power units are in dBm, and $\mathcal{M}$ is the physical uplink shared channel bandwidth of the resource assignment expressed in the number of resource blocks \cite{3gpp.38.213}. In our simulations, we assume that the full bandwidth is used.

\begin{figure}[t]
    \centering
    \includegraphics[width=0.9\textwidth]{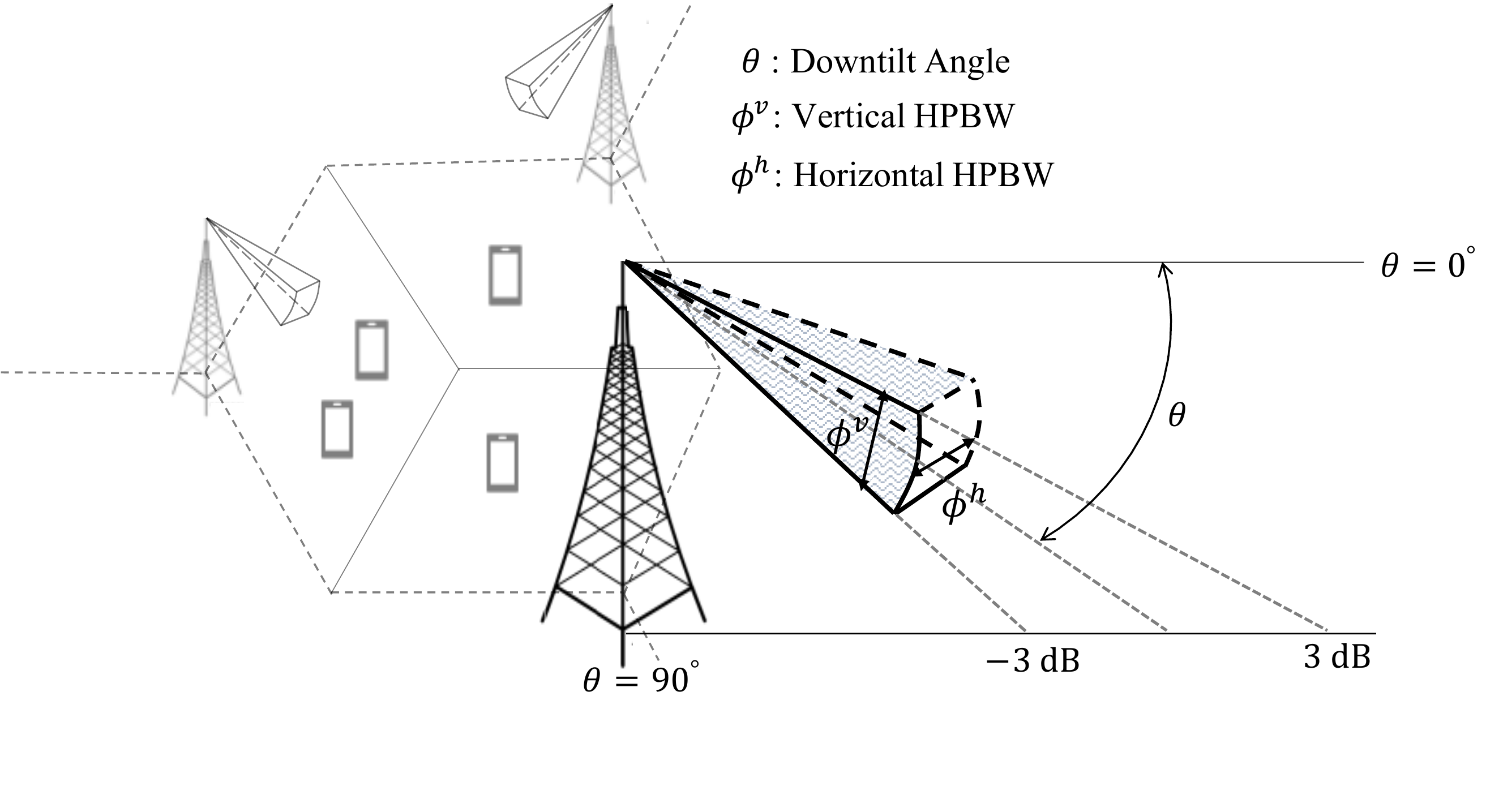}
    \vspace{-0.4cm}
    \caption{Illustration of antenna downtilt angle, vertical half-power beamwidth (HPBW), and horizontal HPBW.}
    \label{fig:beam_new}
    \vspace{-0.2cm}
\end{figure}

Downtilt angle $\theta$, vertical HPBW $\phi^{v}$, and horizontal HPBW $\phi^{h}$ are the parameters of the BS antennas that we attempt to optimize, as seen in Fig. \ref{fig:beam_new}.   A configuration formed by their joint settings over all $M$ cells is represented by $\boldsymbol{x} = \{\boldsymbol{\theta},\boldsymbol{\phi}^{v},\boldsymbol{\phi}^{h}\}$, which we define as a $3M \times 1$ vector (as opposed to a $M \times 3$ matrix) for reasons that we clarify in Sect. \ref{sec:Framework}. The mapping between a given set of antenna settings and the UL and DL SINR of each UE is computed by the Wireless Next-Generation Simulator (WiNGS) developed by AT{\&}T Labs.  We utilize WiNGS to obtain realistic measurements of SINR which could be substituted with actual SINR measurements (and CQI feedback) in real-world implementations. After the parameters of each BS antenna are set, WiNGS simulates the entire network and provides SINR values averaged over both the temporal and spatial domain: specifically, for the RSRP/SINR measurement the channel is sampled every $100$ms over a $1$s time window.  

In summary, our system model corresponds closely to a medium-sized urban cellular 5G NR network with both macrocells and picocells, with all crucial aspects of the system carefully modeled using a state-of-the-art simulator that is used by AT\&T. More of the numerical values and pertinent details of our model and WiNGS are provided in Section \ref{sec:Simulation}.

\subsection{Capacity and Coverage Metrics}
Our \emph{coverage} metric is the (empirical) outage probability $\zeta$.  The outage probability $\zeta$ is defined as the fraction of UEs with SINR below $\mathsf{T}$ dB, i.e., 
\begin{align}\label{eq:outage_prob}
\zeta \triangleq \frac{\sum_{n = 1}^{N}\mathds{1}{\left\{\sinr_{n}(\boldsymbol{x}) < \mathsf{T}\right\}}}{N}, 
\end{align}
where $\sinr_{n}(\boldsymbol{x})$ is the linear SINR of the $n$-th UE for the antenna settings $\boldsymbol{x}$.

Our \emph{rate} metric is average sum-log-rate, which is defined as
\begin{align}
R = \frac{1}{N} \sum_{n=1}^{N} \log \Big(\log\big(1 + \sinr_{n}(\boldsymbol{x})\big)\Big). 
\label{eq:throughput_metric}
\end{align}
One reason for using the \emph{average} sum-log-rate is that the normalization factor $N$ allows the throughput and coverage metrics to have numerical values with approximately the same order or magnitude, which eases both the formulation of a simple optimization problem and the numerical solution of it.  

We choose sum-log-rate as opposed to sum-rate for two reasons.  First, sum-log-rate is a well-accepted rate utility function that balances between sum-rate maximization and fairness across the user distribution.  It is frequently used in real-world cellular systems (e.g. proportionally fair scheduling and load balancing).   Second, sum-log-rate avoids the degenerative solutions that often result from (linear) sum-rate maximization, where just one or a small set of users receive all the resources at the expense of the others.

\subsection{Optimization Problem Formulation} \label{sec:Problem_Formulation}
We now formulate our proposed capacity-coverage optimization problem.  The objective is select the three antenna settings -- $\theta$, $\phi^{v}$, and $\phi^{h}$ -- for each cell that maximizes an arbitrarily weighted combination of capacity and coverage for both the downlink and uplink.

The problem is as follows.
\begin{subequations} \label{eq:optimization_prob_ULDL}
\begin{align}
\max_{\boldsymbol{x}= \{\boldsymbol{\theta},\boldsymbol{\phi}^{v},\boldsymbol{\phi}^{h}\}} {\mathsf{F}_{\boldsymbol{x}}} = (1-\alpha)f&\big(\sinr^{\rm{DL}}(\boldsymbol{x})\big) + \alpha f\big(\sinr^{\rm{UL}}(\boldsymbol{x})\big) \label{eq:objective_function_ULDL}\\
\textrm{  s.t. } \qquad \qquad \qquad \enspace \theta_{m} &\in \Big(\underline{\theta}_{m}, \bar{\theta}_{m}\Big),\\
\phi^{v}_{m} &\in \Big(\underline{\phi}^{v}_{m},\bar{\phi}^{v}_{m}\Big),\\
\phi^{h}_{m} &\in \Big(\underline{\phi}^{h}_{m},\bar{\phi}^{h}_{m}\Big), \enspace m = 1, \dots, M,
\end{align}
\end{subequations}
where
\begin{subequations}
\begin{align}
f\big(\sinr^{\rm{DL}}(\boldsymbol{x})\big) &= \beta^{\rm{DL}} R^{\rm{DL}} - \big(1-\beta^{\rm{DL}}\big)\zeta^{\rm{DL}},\\
f\big(\sinr^{\rm{UL}}(\boldsymbol{x})\big) &= \beta^{\rm{UL}} R^{\rm{UL}} - \big(1-\beta^{\rm{UL}}\big)\zeta^{\rm{UL}},
\end{align}
\end{subequations}

Specifically, $\theta_{m}$ is the downtilt angle, $\phi^{v}_{m}$ is the vertical HPBW, and $\phi^{h}_{m}$ is the horizontal HPBW of $m$-th cell, which yields the vector notation $\boldsymbol{\theta} = \left[\theta_{1}, \dots, \theta_{M}\right] \in \mathbb{R}_{+}^{1 \times M}$, $\boldsymbol{\phi}^{v} = \left[\phi^{v}_{1}, \dots,\phi^{v}_{M} \right] \in \mathbb{R}_{+}^{1 \times M}$, and $\boldsymbol{\phi}^{h} = \left[ \phi^{h}_{1}, \dots, \phi^{h}_{M} \right] \in \mathbb{R}_{+}^{1 \times M}$. The smallest allowed settings are $\underline{\theta}_{m}$, $\underline{\phi}_{m}^{v}$, $\underline{\phi}_{m}^{h}$, while $\bar{\theta}_{m}$ $\bar{\phi}^{v}_{m}$, $\bar{\phi}^{h}_{m}$ are the largest allowed settings.  

The coverage metric $\zeta$ and throughput $R$ are defined as in \eqref{eq:outage_prob} and \eqref{eq:throughput_metric}, respectively, where $\sinr^{\rm{DL}}(\boldsymbol{x}) = \big[\sinr_{1}^{\rm{DL}}(\boldsymbol{x}), \dots, \sinr_{N}^{\rm{DL}}(\boldsymbol{x})\big] \in \mathbb{R}^{1 \times N}$ is a vector of all $N$ users DL SINR's, and similarly for $\sinr^{\rm{UL}}(\boldsymbol{x})$ for the UL SINR values.

The trade-off between downlink and uplink optimization is determined by an \emph{uplink weighting coefficient}, denoted by $\alpha \in [0,1]$.   The trade-off between outage and throughput can be adjusted by a \emph{rate weighting coefficient}, denoted by $\beta \in [0,1]$, which actually consists of two values $\beta^{\rm DL}$ and $\beta^{\rm UL}$, which allows different coverage-capacity tradeoffs in each link.  

The optimization problem \eqref{eq:optimization_prob_ULDL} is nonconvex.  The joint optimization of rate and coverage is a challenging problem due to their conflicting nature, while a multi-cell environment creates coupling between each cell's optimum settings: the optimal configuration of one cell's antenna array depends on the settings of each neighboring cell due to inter-cell interference.   The formulated problem has three parameters to be configured for each cell, and hence there are a total of $3M$ optimization parameters each of which is continuous over the specified range.  The search space considering all cells is exponential in $M$, rendering an exhaustive search for the best settings impossible for moderate values of $M$, even if the search space is discretized.  

Thus, computationally efficient methods that approach the unknowable optimum solution are desirable.   It should be noted that calculating (or measuring) all the SINR values is quite costly and time consuming, whether by simulation or via real-world measurements.   Hence, each iteration of the algorithm or computation of \eqref{eq:optimization_prob_ULDL} occurs at a considerable cost.   Sample-efficient algorithms are thus highly desirable. 

%
%
\vspace{-0.2cm}
\section{Proposed Learning Framework} \label{sec:Framework}
The enumerated challenges of the site-specific antenna tuning problem motivate us to adopt a data-driven and sample-efficient learning framework, and we propose a methodology leveraging desirable features of the machine learning-based black-box optimization technique Bayesian optimization (BO) and a metaheuristic search algorithm differential evolution (DE) to approximately solve the non-convex antenna optimization problem formulated in Section \ref{sec:Problem_Formulation}. We are inspired by BO since it is well-suited for solving such expensive-to-evaluate black-box optimizations with a limited computational budget \cite{jin2021data}, and it provides an accurate model that enables the algorithm to eliminate the candidates which are not promising and to spend its resources only on the most promising candidates. The more candidate can thus be evaluated with less computational resources. 

\begin{figure}[h]
    \centering
    \includegraphics[width = 0.8\textwidth]{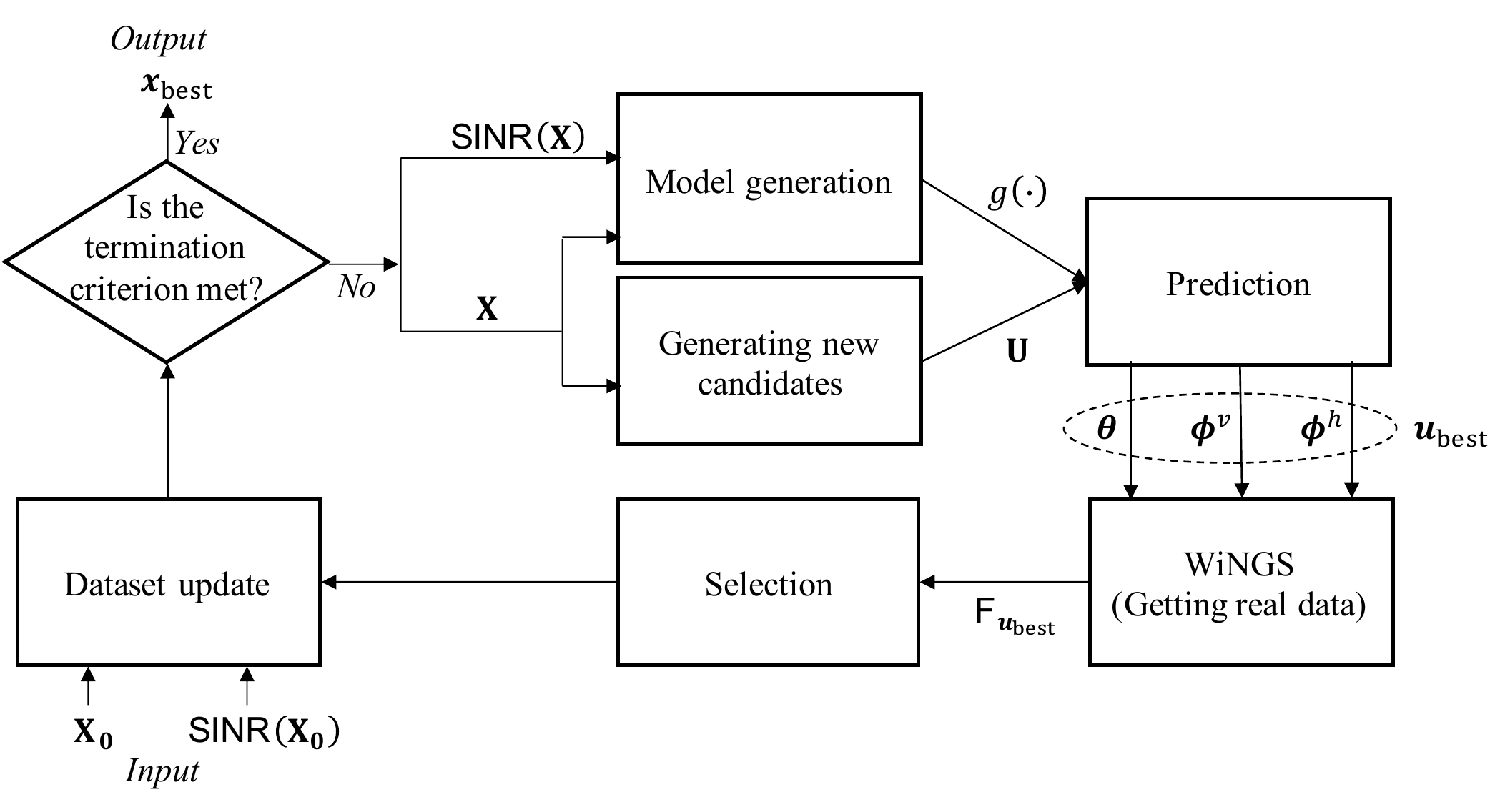}
    \caption{The flowchart of the proposed algorithm, where $\mathbf{X}_{0}$ is the randomly initialized population matrix at Step \ref{step:initialization}.}
    \label{fig:flowchart}
    \vspace{-0.4cm}
\end{figure}

However, BO has some drawbacks that hinder its practical application in a real-world environment. For example, it has a cubic complexity and requires solving a non-convex problem with
increasing computational cost while deciding on the next sampling point, making its cost prohibitive as the number of iterations (i.e., training data size) increases \cite{shahriari2015taking}. This motivates us to adopt a different search algorithm to combine with the model building part of BO, and we opt for the evolutionary search framework, considering its strong search capabilities (i.e., exploring multiple areas of the search space simultaneously and being able to find solutions in high-dimensional search spaces \cite{miikkulainen2021creative}) and the constant time it takes to generate new candidate solutions. This section details the steps of our proposed learning framework whose flowchart is depicted in Fig. \ref{fig:flowchart}.
\subsection{Generating New Candidates} \label{sec:New_Candidate_Gen}
The algorithm starts by randomly initializing a population matrix, $\mathbf{X} \in \mathbb{R}_{+}^{S \times 3M}$, with $S$ individual vectors (i.e., antenna configurations) and calculating its real objective value, $\mathsf{F}_{\mathbf{X}}$. The $i$-th individual vector in $\mathbf{X}$ is
\begin{equation}
	    \begin{aligned}
	        \boldsymbol{x}_{i} = [\boldsymbol{\theta}_{i},\boldsymbol{\phi}^{v}_{i}, \boldsymbol{\phi}^{h}_{i}],
	    \end{aligned}
\end{equation}
where $\boldsymbol{\theta}_{i} = [\theta_{i,1}, \dots, \theta_{i,M}]$, $\boldsymbol{\phi}^{v}_{i} = [\phi^{v}_{i,1}, \dots, \phi^{v}_{i,M}]$, and $\boldsymbol{\phi}^{h}_{i} = [\phi^{h}_{i,1}, \dots, \phi^{h}_{i,M}]$ denote the $i$-th instance of the vectors $\boldsymbol{\theta}, \boldsymbol{\phi}^{v},$ and $\boldsymbol{\phi}^{h}$, respectively, and $\mathbf{X} = [\boldsymbol{x}_{1}^{T}, \dots, \boldsymbol{x}_{S}^{T}]^{T}$. The next step is generating and evaluating new candidates, which is performed by a modified DE framework. DE is a population-based evolutionary algorithm. As with other evolutionary algorithms, it reaches better solutions through mutation \eqref{eq:V}, crossover \eqref{eq:U}, and the selection strategy that keeps the member with the better objective function value. Hence, once the population is initialized with the matrix $\mathbf{X}$, a mutant vector is generated from each individual in $\mathbf{X}$ using one of the various kinds of DE mutation strategies which are denoted as DE/\textit{a}/\textit{b}, where \textit{a} represents the vector to be mutated and \textit{b} is the number of difference vectors used \cite{price2006differential}. Among these strategies, we adopt DE/current-to-best/1 mutation scheme, which generates a mutant vector for the $i$-th individual as follows,
\begin{equation} \label{eq:V}
\boldsymbol{v}_{i} = \boldsymbol{x}_{i} + F\cdot(\boldsymbol{x}_{\text{best}} - \boldsymbol{x}_{i}) + F\cdot(\boldsymbol{x}_{r_{1}} - \boldsymbol{x}_{r_{2}}),
\end{equation}
where the scale factor $F$ is a positive real number that controls the population evolving rate and is usually less than $1$. The vector $\boldsymbol{x}_{\text{best}}$ is the individual with the best objective function value in the current population, and $\boldsymbol{x}_{r_{1}}$ and $\boldsymbol{x}_{r_{2}}$ are the randomly chosen individuals. The motivation for choosing this mutation strategy is to reach a compromise between exploitation and exploration \cite{qiang2016tuning}. In the next step, a trial vector $\boldsymbol{u}_{i} = [u_{i,1}, \dots, u_{i,3M}]$ is created for each mutant vector $\boldsymbol{v}_{i} = [v_{i,1}, \dots, v_{i,3M}]$ by carrying out the crossover operator as
\begin{equation} \label{eq:U}
u_{i,j} =\begin{cases} 
v_{i,j}, & \text{with probability } p_{c} \\
x_{i,j}, & \text{otherwise} \\
\end{cases},
\end{equation}
where $p_{c}$ is the crossover probability and determines the fraction of the trial vector that comes from the mutant vector \cite{5601760}. Finally, in DE, a selection is made between each individual in the current population ($\boldsymbol{x}_{i}$, $\forall i$), and the corresponding trial vector, $\boldsymbol{u}_{i}$, based on the calculated true objective function values, determining the population of the next iteration. The modification is made in this final step, where DE calculates the real objective function value of all individuals in the population at each iteration. This makes DE a computationally expensive and time-consuming algorithm. However, in the proposed hybrid algorithm, this part is replaced by the model building and prediction part of the BO with additional modifications (Section \ref{sec:Model_Gen}). The computationally inexpensive surrogate model is thus substituted for time-consuming real simulations or explorations.
\vspace{-0.2cm}
\subsection{Model Generation and Prediction} \label{sec:Model_Gen}
The two major components of BO are a Bayesian statistical model (i.e., surrogate function) to model the objective function, and an acquisition function to decide on the next sample \cite{frazier2018bayesian}. We employ only the first component in the proposed hybrid algorithm and choose a stochastic model Gaussian process (GP) as the surrogate function. We further differentiate from conventional BO by modeling UE SINRs independently with different input parameters, and hence having multiple GP models at each iteration. This enables us to have more accurate models and perform modeling in parallel, and thus accommodate a large number of UEs while preserving computational efficiency. Hence, after the trial population $\mathbf{U} = [\boldsymbol{u}_{1}^{T}, \dots, \boldsymbol{u}_{S}^{T}]^{T}$ is created by \eqref{eq:U}, the uplink and downlink SINR models are generated by Gaussian process regression (GPR) for each UE using the current population $\mathbf{X}$. Then, for each individual vector in the trial population ($\boldsymbol{u}_{i}$, $\forall i$), the uplink and downlink SINRs of each UE are predicted. GPR is an interpolation method ruled by prior covariances \cite{jin2021data}. A GPR model, $g(\boldsymbol{x})$, is fully specified by its mean function $\mu(\boldsymbol{x})$ and kernel function $k(\boldsymbol{x}, \boldsymbol{x}')$: $g(\boldsymbol{x}) \sim \mathcal{GP}(\mu(\boldsymbol{x}), k(\boldsymbol{x},\boldsymbol{x}'))$. To make predictions about unseen test cases, GPR utilizes the posterior. As the current population $\mathbf{X}$ is a matrix of training inputs and the trial population $\mathbf{U}$ is a matrix of test inputs, the conditional posterior for the test points is 
\begin{equation}
P(\boldsymbol{g}_{\mathbf{U}}|\mathbf{X},\boldsymbol{y},\mathbf{U}) \sim \mathcal{N}(\boldsymbol{\bar{g}}_{\mathbf{U}}, \text{cov }\boldsymbol{g}_{\mathbf{U}}), 
\end{equation}
where 
\begin{equation}\label{eq:mean_for_test_point}
\boldsymbol{\bar{g}}_{\mathbf{U}} = K(\mathbf{U},\mathbf{X})[K(\mathbf{X},\mathbf{X})+\sigma_{n}^{2}I]^{-1}\boldsymbol{y}, 
\end{equation}
\begin{equation}\label{eq:cov_for_test_point}
\text{cov } \boldsymbol{g}_{\mathbf{U}} = K(\mathbf{U},\mathbf{U})-K(\mathbf{U},\mathbf{X})[K(\mathbf{X},\mathbf{X})+ \sigma_{n}^{2}I]^{-1}K(\mathbf{X},\mathbf{U}), 
\end{equation}
$\boldsymbol{g}_{\mathbf{U}} = \left[g(\boldsymbol{u}_{1}), \dots, g(\boldsymbol{u}_{S})\right]$, $\boldsymbol{y} = \left[\sinr_{\mathbf{X}}^{\mathrm{DL}} \enspace \sinr_{\mathbf{X}}^{\mathrm{UL}}\right]$, and $\sigma_{n}^{2}$ is an independent noise variance.  $\sinr_{\mathbf{X}}^{\mathrm{DL}}$ and $\sinr_{\mathbf{X}}^{\mathrm{UL}} \in \mathbb{R}^{S \times N}$ are the downlink and uplink UE SINR values, respectively, corresponding to each antenna configuration in $\mathbf{X}$, and $K(.,.)$ denotes the covariance matrix. For a more detailed explanation of GPs, please see \cite{rasmussen2006}.
\vspace{-0.2cm}
\subsection{Selection} \label{sec:Selection}
After UE SINR values, $\hat{\sinr}^{\rm{DL}}(\mathbf{U})$ and $\hat{\sinr}^{\rm{UL}}(\mathbf{U})$, are predicted with the GPR model, the objective function value of the trial population, $\mathsf{F}_{\mathbf{U}}$, is calculated using these predictions. The individual with the best estimated objective function value is then chosen among the trial population, and only that individual’s real objective function value, $\mathsf{F}_{\boldsymbol{u}_{\text{best}}}$, is obtained from the simulator. If this value is better than the worst objective function value in the current population, $\mathsf{F}_{\boldsymbol{x}_{\text{worst}}} = \min\left(\mathsf{F}_{\mathbf{X}}\right)$, the new candidate solution, $\boldsymbol{u}_{\text{best}}$, is substituted for that worst individual, $\boldsymbol{x}_{\text{worst}}$. The training dataset is thus updated. In each iteration, we have a population $\mathbf{X}$ with $S$ individual solutions, including at most one new solution. This  makes  the  computational  cost  of  GP  modeling constant by keeping the training data size fixed. As this process is repeated, the current population progresses together to a better region, and when the termination criterion is satisfied, the best individual in the current population, $\boldsymbol{x}_{\text{best}}$, is chosen as the desired solution.
\vspace{-0.2cm}
\subsection{Neighborhood Approach}
A caveat of using GP is that it restricts the dimension of the problem as it is best suited for optimization over continuous domains with about twenty or fewer decision variables (i.e., input parameters) \cite{frazier2018bayesian}. It has also been investigated for medium-scale problems with 20-50 decision variables, concluding that it can still be an effective approach in that range \cite{liu2013gaussian}. However, it is not suitable for higher dimensional problems. In the defined problem, \eqref{eq:optimization_prob_ULDL}, we have $3M$ input parameters, corresponding to $M$ downtilt angles, $M$ vertical HPBWs, and $M$ horizontal HPBWs, where $M$ is the number of total cells in the network. Hence, this dimensionality constraint restricts the network size. To remove this restriction, we propose to define a \textit{neighborhood} for each UE, which includes cells with a large measured RSRP. Considering the fact that the cells with low RSRP do not have a significant effect on the UE SINR, we only consider the neighboring cells of a UE while modeling its SINR. This significantly reduces the dimension of the problem and helps GP have an accurate model, which enables the algorithm to be successfully scaled to much larger networks. 

The neighborhoods are defined once at the beginning of the proposed algorithm by initializing the parameters of all cells to the specific values -- $\boldsymbol{\theta}_{*}, \boldsymbol{\phi}^{v}_{*}, \boldsymbol{\phi}^{h}_{*}$ -- and neighboring cells are then chosen according to calculated uplink and downlink RSRP values. Namely, for a $\mathcal{N}$-sized neighborhood, $\mathcal{N}$ cells with the highest RSRP values are selected for each UE. In our simulations, we validate our neighborhood approach by ensuring that for each UE, the strongest $\gamma\mathcal{N}$ interferers are captured in its neighborhood with a high probability as the parameter values change, where the $\gamma$ is a fraction between $0.6$ and $1$.
\begin{algorithm}
\begin{algorithmic}[1]
    \Require $\mathbf{X} = \left[ \boldsymbol{x}_{1}^{T}, \dots, \boldsymbol{x}_{S}^{T} \right]^{T}, \sinr(\mathbf{X}), \mathsf{F}_{\mathbf{X}}$  \label{step:initialization}
    \For {$\text{iter} = 1, 2, \dots, N_{\text{iter}}$}
    \State $\boldsymbol{v}_{i} = \boldsymbol{x}_{i} + F\cdot(\boldsymbol{x}_{\text{best}} - \boldsymbol{x}_{i}) + F\cdot(\boldsymbol{x}_{r_{1}} - \boldsymbol{x}_{r_{2}}), \forall i$ \label{step: Mutation} 
    \State $ u_{i,j} = \begin{cases} 
v_{i,j}, & \text{with probability } p_{c} \enspace  \\
x_{i,j}, & \text{otherwise}
\end{cases}, \boldsymbol{u}_{i} = [u_{i,1}, \dots, u_{i,3M}], \mathbf{U} = [\boldsymbol{u}_{1}^{T}, \dots, \boldsymbol{u}_{S}^{T}]^{T}$ \label{step:Recombination}
        \For {$n = 1, \dots, N$}
 	    \State Construct GP models for uplink and downlink SINR of $n$-th UE using the antenna parameters of its neighboring cells. \label{step:GP_model_gen}
 	    \State Predict $\hat{\sinr}^{\rm{DL}}_{n}(\mathbf{U})$ and $\hat{\sinr}^{\rm{UL}}_{n}(\mathbf{U})$ with the created GP models.
 	    \EndFor
 	    \State $\mathsf{F}_{\mathbf{U}} = (1-\alpha)f\big(\hat{\sinr}^{\rm{DL}}(\mathbf{X})\big) + \alpha f\big(\hat{\sinr}^{\rm{UL}}(\mathbf{X})\big)$ 
	    \State $\boldsymbol{u}_{\text{best}} = \arg\max_{\boldsymbol{u}} \mathsf{F}_{\mathbf{U}}$ \label{step:selection_1}
    	\State Observe $\mathsf{F}_{\boldsymbol{u}_{\text{best}}}$ \label{step:selection_2}
	    \If {$\mathsf{F}_{\boldsymbol{u}_{\text{best}}} \geq \mathsf{F}_{\boldsymbol{x}_{\text{worst}}} = \min\left(\mathsf{F}_{\mathbf{X}}\right)$} \label{step:selection_3}
	    \State $\boldsymbol{x}_{\text{worst}} \leftarrow \boldsymbol{u}_{\text{best}}$ \label{step:selection_4}
	    \EndIf
	\EndFor
\Ensure $\boldsymbol{x}_{\text{best}} = \arg\max_{\boldsymbol{x}} \mathsf{F}_{\mathbf{X}}$ \label{step:output}
 \caption{Sample-Efficient Learning Algorithm}
 \label{Algorithm}
\end{algorithmic}
\end{algorithm}

Overall, our proposed sample-efficient learning algorithm, whose steps are summarized in Algorithm \ref{Algorithm}, begins by defining a neighborhood for each UE. Then, it follows the steps explained in Section \ref{sec:New_Candidate_Gen} to \ref{sec:Selection}. Notice that with the involvement of the neighborhood approach, the input vector of the $n$-th UE's GPR model consists of only the parameters of its neighboring cells. Hence, each UE has a different input parameter population to define a model for its SINR.
\begin{enumerate}
    \item \textbf{Average sum-log-rate}, $R$, is defined in \eqref{eq:throughput_metric}. It is a throughput metric used in the objective function, and as a performance metric, it measures the throughput improvement of the proposed algorithm compared to the 3GPP default settings and other comparative algorithms.
    \item \textbf{Outage probability} is the fraction of UEs with an SINR value less than the threshold $\mathsf{T}$ dB and is defined in \eqref{eq:outage_prob}. This metric measures the coverage improvement that the proposed algorithm achieves compared to the 3GPP default settings and other comparative methods.

    \item \textbf{UE SINR} values are derived from WiNGS after the antenna parameter values of the cells are optimized, and their empirical CDF plots are compared with the 3GPP default settings and other comparative algorithms. In addition to measuring the UE SINR improvement, this metric allows us to observe the fairness of the algorithm by tracking which SINR regions the improvements are in.
\end{enumerate}
\vspace{-0.4cm}
%
\section{Simulation Details} \label{sec:Simulation}
We perform our experiments on a state-of-the-art Wireless Next-Generation Simulator (WiNGS) developed by AT{\&}T Labs. WiNGS is used within AT{\&}T for developing and evaluating advanced air interface and radio access network features across a range of realistic deployment scenarios based on both statistical modeling tools and real-world network data input. This event-driven, modular, fully dynamic system level simulator (SLS) closely models the air interface functionality and operations -- including PHY, MAC, RLC, PDCP, SDAP, and RRC layers -- of the 5G New Radio (NR) radio access network protocol stack. The wireless channels for the access links are generated using 3GPP-defined statistical models with both long-term line-of-sight and non-line-of-sight path loss, shadowing, and short-term fading effects. BS deployments can be modeled based on a fixed grid or random heterogeneous layouts and even based on real-world deployment data, including modeling antenna array geometry on a per-site basis.
\begin{figure}[ht]
    \hspace{-0.7cm}
    \begin{minipage}[b]{0.52\linewidth}
    \centering
    \includegraphics[width=\textwidth]{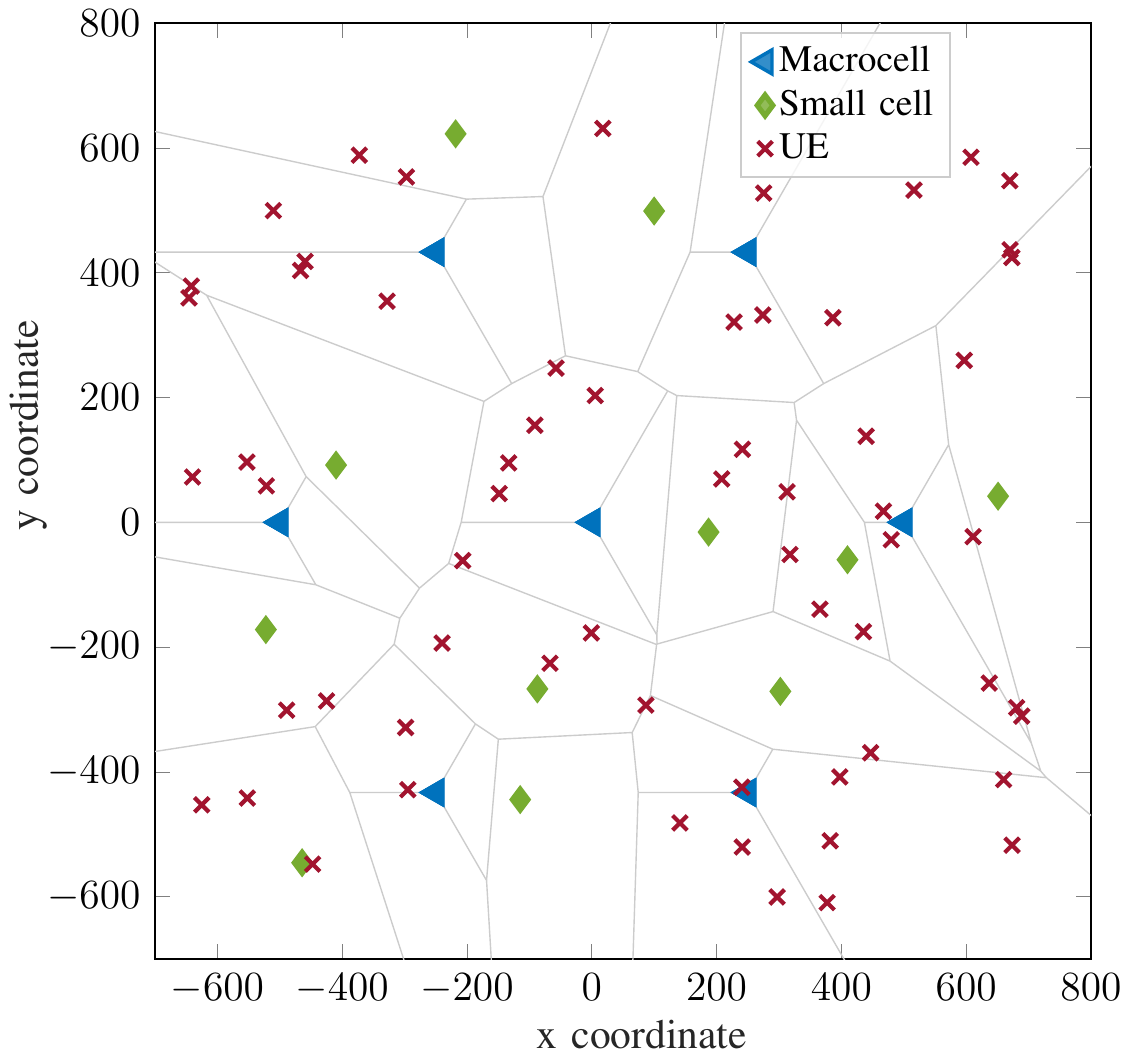}
     \subcaption{Layout 1}
    \end{minipage}%
    \hspace{-0.2cm}
    \begin{minipage}[b]{0.52\linewidth}
    \includegraphics[width=\textwidth]{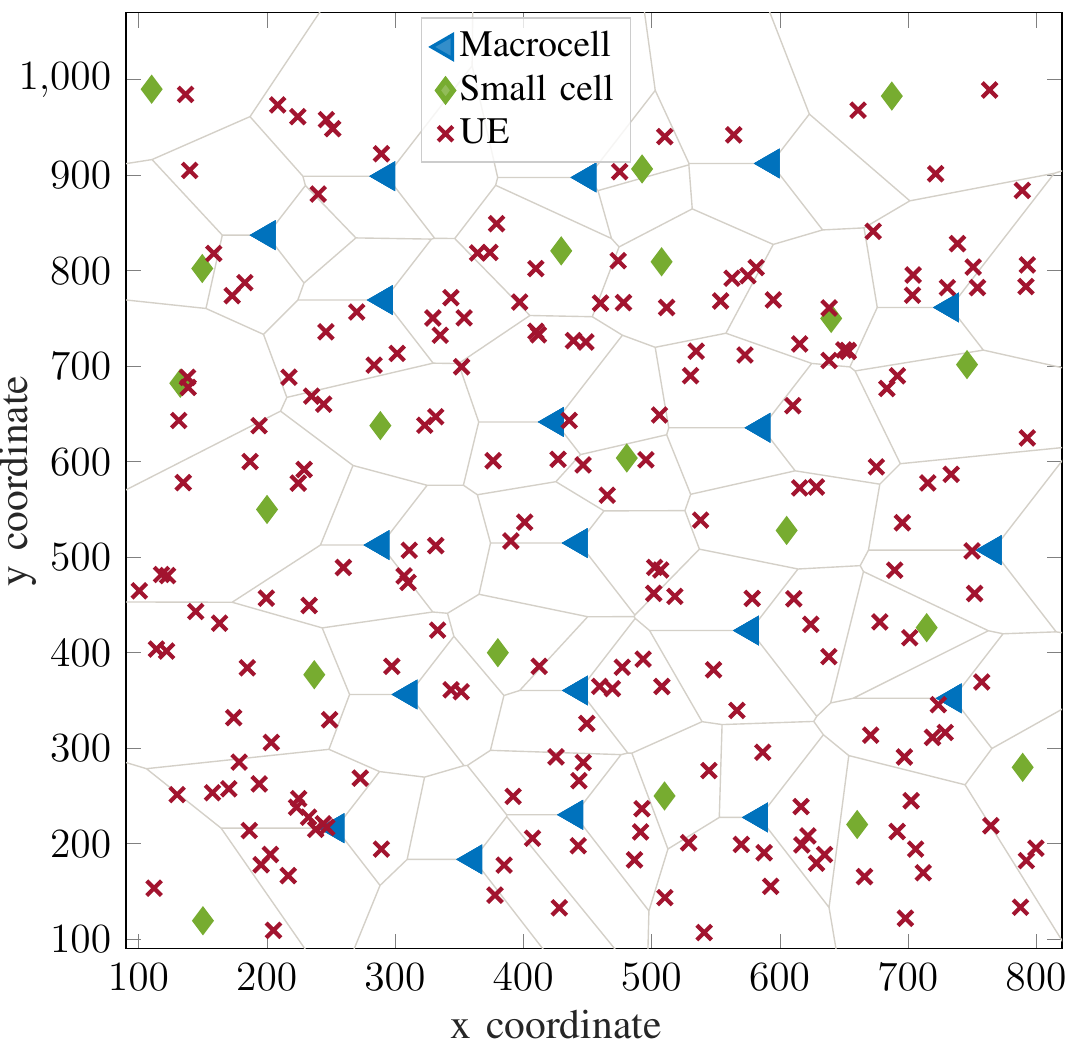}
     \subcaption{Layout 2}
     \label{fig:Layout2}
     \end{minipage}
     \vspace{-0.6cm}
    \caption{Two HetNet layouts with uniformly distributed UEs.}
    \label{fig:hetnet_layouts}
    \vspace{-0.7cm}
\end{figure}

As a dynamic SLS, WiNGS generates a variety of per-user metrics including RSRP, L1 channel state information, and SINR measurements by taking into account inter-cell interference, which are used in the link adaptation and resource allocation blocks for modulation coding scheme and transport block size selection every scheduling interval. In addition, practical digital and analog codebooks are used to support cell-specific and user-specific beamforming.

On this high-fidelity simulator, we consider two outdoor HetNet deployments to verify the proposed framework experimentally and show its scalability to large network sizes. Layout 1 is an hexagonal heterogeneous layout with a total of $M = 32$ cells (7 macrocells with three sector antennas and $11$ small cells) and $N = 62$ UEs which are uniformly distributed to different locations. Layout 2 is based on real-world deployment data, and it comprises 200 uniformly distributed UEs and a total of $M = 77$ cells (19 macrocells with three sector antennas and 20 uniformly distributed small cells). The macrocells have a height of $25$m, while small cells of $10$m and UEs are of $1.5$m. The distributions of macrocells, small cells, and UEs in the two HetNet deployments are shown in Fig. \ref{fig:hetnet_layouts}. 

We compare our proposed algorithm with five different baselines in two different simulation environments. Three of these baselines -- 3GPP default settings, random search, and conventional Bayesian optimization (BO) with expected
improvement -- are implemented in the described simulation environment (i.e., WiNGS), and the results for the other two -- the deep deterministic policy gradient algorithm (DDPG) and BO with q-expected hypervolume improvement \cite{daulton2020differentiable} -- are obtained from the work \cite{dreifuerst2021optimizing} in which a MATLAB tool suite called QuaDRiGa \cite{burkhardt2014quadriga} is used. Regarding the implemented three baselines, in the 3GPP default settings, a single configuration ($\theta = 12^{\circ}, \phi^{v} = 10^{\circ},$ and $\phi^{h} = 70^{\circ}$) that is the same for all cells is chosen through an extensive search, as in 3GPP. Random search draws random samples from the search space and keeps the best configuration at each iteration during optimization. BO and the proposed algorithm use the same kernel function and hyperparameter initialization, and a population with the same number of individuals is randomly initialized at the beginning of random search, BO, and the proposed algorithm. 
The simulations start by initializing the tilt angles, vertical HPBW and horizontal HPBW of each cell's antenna arrays to $\theta_{*} = 12^{\circ}$, $\phi^{v}_{*} = 10^{\circ}$ and $\phi^{h}_{*} = 70^{\circ}$, respectively. A neighborhood for each UE is then determined by choosing $\mathcal{N}=8$ and $\mathcal{N}=10$ neighboring cells whose RSRP values are maximum in Layout 1 and 2, respectively. Throughout the simulation, the parameters of these neighboring cells are used for modeling the SINR of the associated UE. After a neighborhood for each UE is determined, initialization is performed by randomly choosing $S = 200$ antenna parameter configurations for each cell. 
\vspace{-0.2cm}
\begin{table}[h]
\caption{Simulation Parameters}
\begin{minipage}{0.5\linewidth}
\vspace{-0.2cm}
\centering
\resizebox{\textwidth}{!}{%
\begin{tabular}{ |p{3.55cm}||p{1.25cm}|>{\centering\arraybackslash}p{2.5cm}|  }
 \hline
 \textbf{Optimization Parameters}& &\textbf{Values}\\
 \hline\hline
 \rowcolor{gray!10}
 Crossover probability, $p_{c}$&  &$0.8$\\
 Scale factor, $F$&   &$0.7$\\
 \rowcolor{gray!10}
 Population size, $S$&   &$200$\\
 Neighborhood size, $\mathcal{N}$&\makecell{Layout 1 \\ Layout 2}&\makecell{$8$ \\ $10$}\\
 \rowcolor{gray!10}
 Threshold, $\mathsf{T}$& &$0$ dB, $10\%$ $\sinr$ \\
 Total iteration number, $N_{\text{iter}}$& &$1000$ \\
 \rowcolor{gray!10}
 \makecell[l]{Sum-log-rate and outage \\ probability trade-off, $\beta$}& &$[0.2, 0.5, 0.8, 1]$ \\
 UL weight (vs. DL), $\alpha$ & &$[0.2, 0.5, 0.8]$ \\
  \rowcolor{gray!10}
 $\{\underline{\theta}, \underline{\phi}^{v}, \underline{\phi}^{h}\}$& &$\{0^{\circ}, 0^{\circ}, 5^{\circ}\}$\\
 $\{\bar{\theta},\bar{\phi}^{v}, \bar{\phi}^{h}\}$& &$\{25^{\circ}, 65^{\circ}, 100^{\circ}\}$\\
  \rowcolor{gray!10}
 $\{\boldsymbol{\theta}^{*}, \boldsymbol{\phi}^{v}_{*}, \boldsymbol{\phi}^{h}_{*}\}$& &$\{12^{\circ}, 10^{\circ}, 70^{\circ}\}\boldsymbol{1}_{M}$\\
 \hline
\end{tabular}%
}
\end{minipage}
\begin{minipage}{0.5\linewidth}
\vspace{-0.2cm}
\centering
\resizebox{\textwidth}{!}{%
\begin{tabular}{ |p{4.25cm}||p{1.25cm}|>{\centering\arraybackslash}p{1.8cm}|  }
 \hline
 \textbf{System Parameters}& &\textbf{Values}\\
 \hline\hline
  \rowcolor{gray!10}
     Carrier frequency&  &$2$ GHz   \\ 
     Bandwidth&    &$10$ MHz   \\ 
      \rowcolor{gray!10}
     Subcarrier spacing&  &$15$ kHz   \\ 
     Number of users, $N$&\makecell{Layout 1 \\ Layout 2}&\makecell{$62$ \\  $200$}\\
 \rowcolor{gray!10}
     Number of cells, $M$&\makecell{Layout 1 \\ Layout 2}&\makecell{$32$ \\ $77$}\\
     Macrocell transmit power&    &$43$ dBm  \\
      \rowcolor{gray!10}
     Small cell transmit power&  &$30$ dBm  \\
     Maximum uplink transmit power&   &$23$ dBm  \\
      \rowcolor{gray!10}
     Macrocell antenna height& &$25$m  \\
     Small cell antenna height&   &$10$m   \\
      \rowcolor{gray!10}
     User height&  &$1.5$m  \\ 
  \hline
\end{tabular}%
}
\end{minipage}
\label{table:sim_params}
\end{table}

In all subsequent simulations, the Matérn $5/2$ ARD kernel function is used for GP. It is one of the most common kernel classes used in GP and a popular choice in BO due to its flexibility in smoothness. The Matérn covariance between two data points, $x_{i}$ and $x_{j}$ is
\begin{equation} \label{kernel}
\small
k(x_{i},x_{j}) = \frac{2^{1-\nu}}{\Gamma(\nu)}\Big(\frac{\sqrt{2\nu}}{l}d(x_{i},x_{j})\Big)^{\nu}K_{\nu}\Big(\frac{\sqrt{2\nu}}{l}d(x_{i},x_{j})\Big),
\end{equation}
where $K_{\nu}$ is a modified Bessel function, and $\nu$ and $l$ are the positive hyperparameters of the kernel function. The value $\nu$ controls the smoothness of the function while $l$ is a length scale parameter. The optimization of these parameters follows a maximum likelihood estimation method \cite{rasmussen2006}, and the gpml package \cite{rasmussen2010gaussian} is used in the implementation of GP.

Control parameters of DE -- crossover probability $p_{c}$ and scale factor $F$ -- are chosen as $0.8$ and $0.7$, respectively. These values are chosen experimentally. In the literature, a reasonable value for $F$ is usually between $0.4$ and $1$ and $p_{c}$ is between $0.3$ to $0.9$, where the higher values of $p_{c}$ speed up convergence. Overall, these values depend on the objective function chosen and the problem. The key simulation parameters are summarized in Table \ref{table:sim_params}.
\vspace{-0.2cm}
%
\section{Results and Discussion} \label{sec:Results_Discussion}
The performance of the proposed algorithm is evaluated in the environments described in Section \ref{sec:Simulation} for downlink-only (i.e., $\alpha  = 0$), uplink-only (i.e., $\alpha  = 1$), and uplink and downlink joint directions of a cellular system. The results showing the performance of the proposed algorithm compared to other baseline algorithms are presented in Fig. \ref{fig:DLonly_performance_layout1} - \ref{fig:DLonly_UESINR} for Layout 1 and 2. Downlink-only optimization is used, and downlink performance metrics are plotted in these figures. The performance of the uplink and downlink joint optimization compared to the downlink-only and uplink-only optimization are then presented in Fig. \ref{fig:alpha05_UESINR} and \ref{fig:alpha02_UESINR} for Layout 1. The results are obtained by implementing and evaluating the proposed algorithm and the comparative baselines on WiNGS. They are calculated over 5 realizations of each layout and plotted with the results using 3GPP default settings, where natural logarithm is used in the average sum-log-rate plots. Finally, the comparison results of our proposed algorithm with DDPG and BO with q-expected hypervolume improvement are presented in Fig. \ref{fig:complexity_analysis}.
%
\vspace{-0.4cm}
\subsection{Performance Comparisons with Baseline Algorithms} \label{sec:Downlink_opt}
In this section, we compare the numerical results of our approach with other baseline algorithms where the values $\alpha$, $\beta^{\rm{DL}}$, and $\mathsf{T}$ in the problem \eqref{eq:optimization_prob_ULDL} are set as $0$, $0.5$, and $0$ dB, respectively.

\begin{figure}[h]
    \hspace{-0.5cm}
    \begin{minipage}[b]{0.5\linewidth}
       \vspace{-0.4cm}
        \centering
        \includegraphics[width=\textwidth]{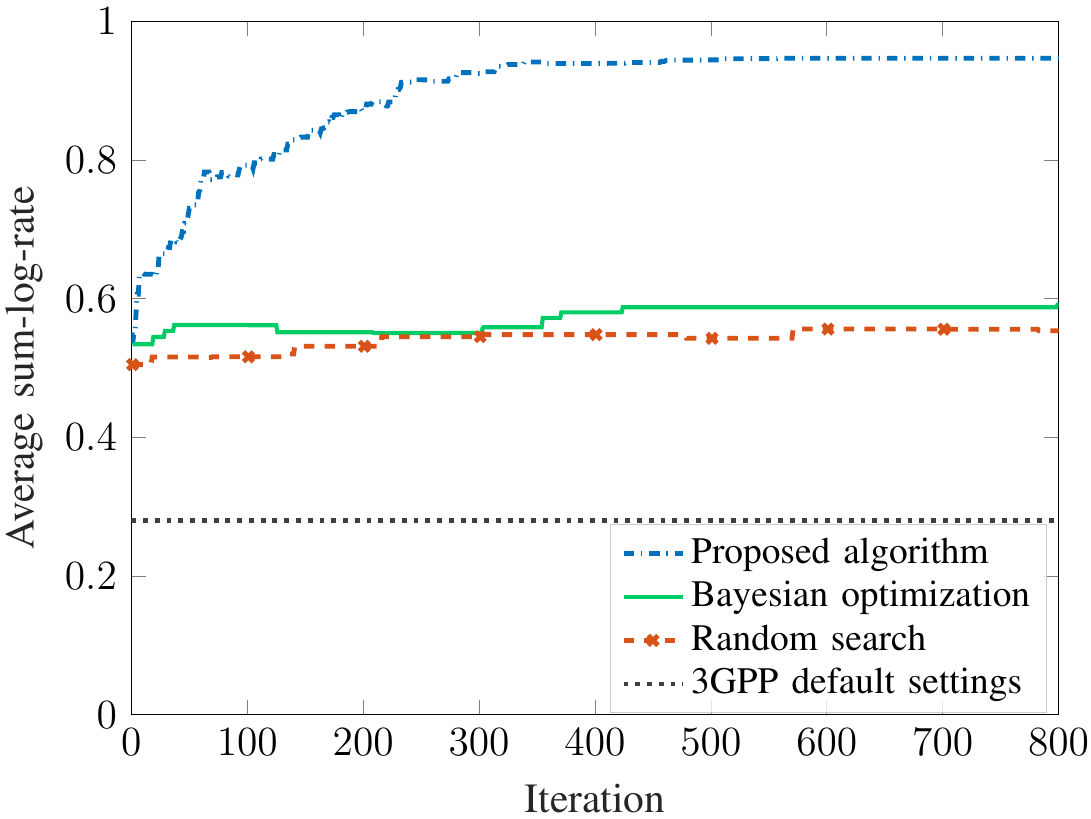}
    \subcaption{Average sum-log-rate vs. iteration}
    \label{fig:R}
    \end{minipage}%
    \enspace
    \begin{minipage}[b]{0.5\linewidth}
       \vspace{-0.4cm}
        \centering
        \includegraphics[width=\textwidth]{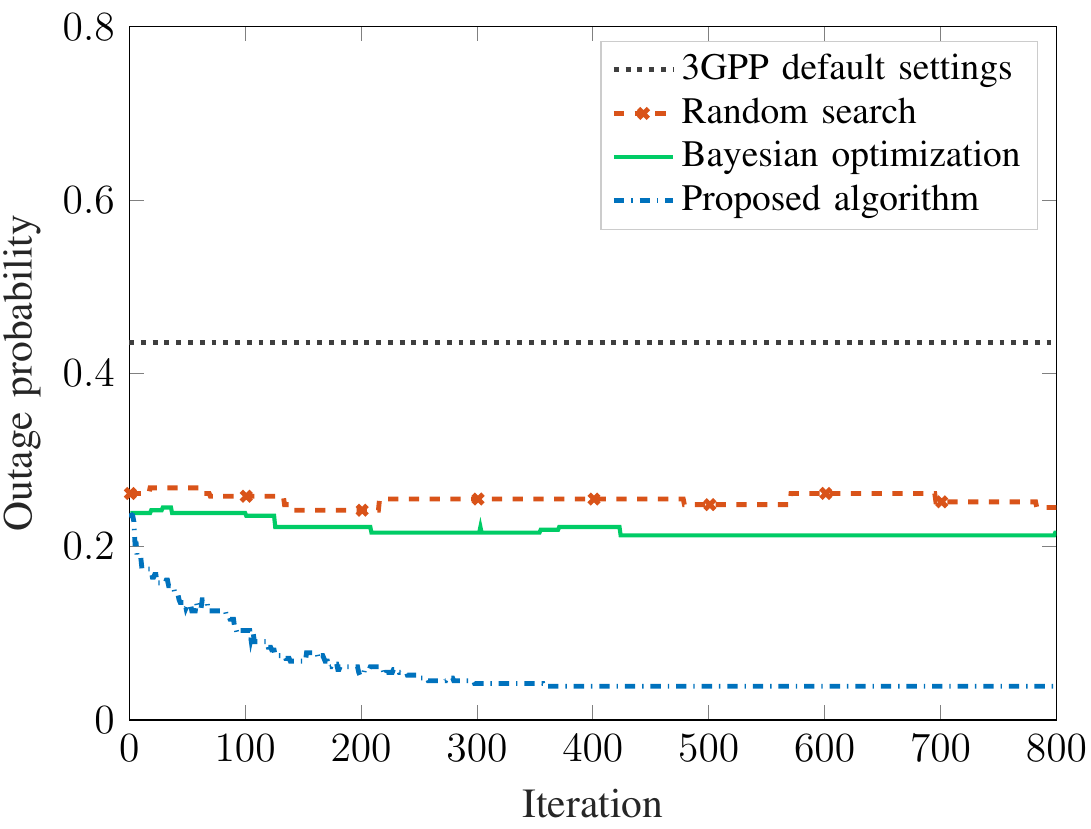}
    \subcaption{Outage probability vs. iteration}
    \label{fig:zeta}
    \end{minipage}
    \vspace{-0.2cm}
\caption{Comparison of the proposed algorithm, 3GPP default settings, random search, and Bayesian optimization in terms of average sum-log-rate and outage probability values versus iteration in Layout 1.}
\label{fig:DLonly_performance_layout1}
\vspace{-0.4cm}
\end{figure}
\begin{figure}[h]
    \hspace{-0.6cm}
    \begin{minipage}[b]{0.5\linewidth}
    \vspace{-0.2cm}
    \centering
    \includegraphics[width=\textwidth]{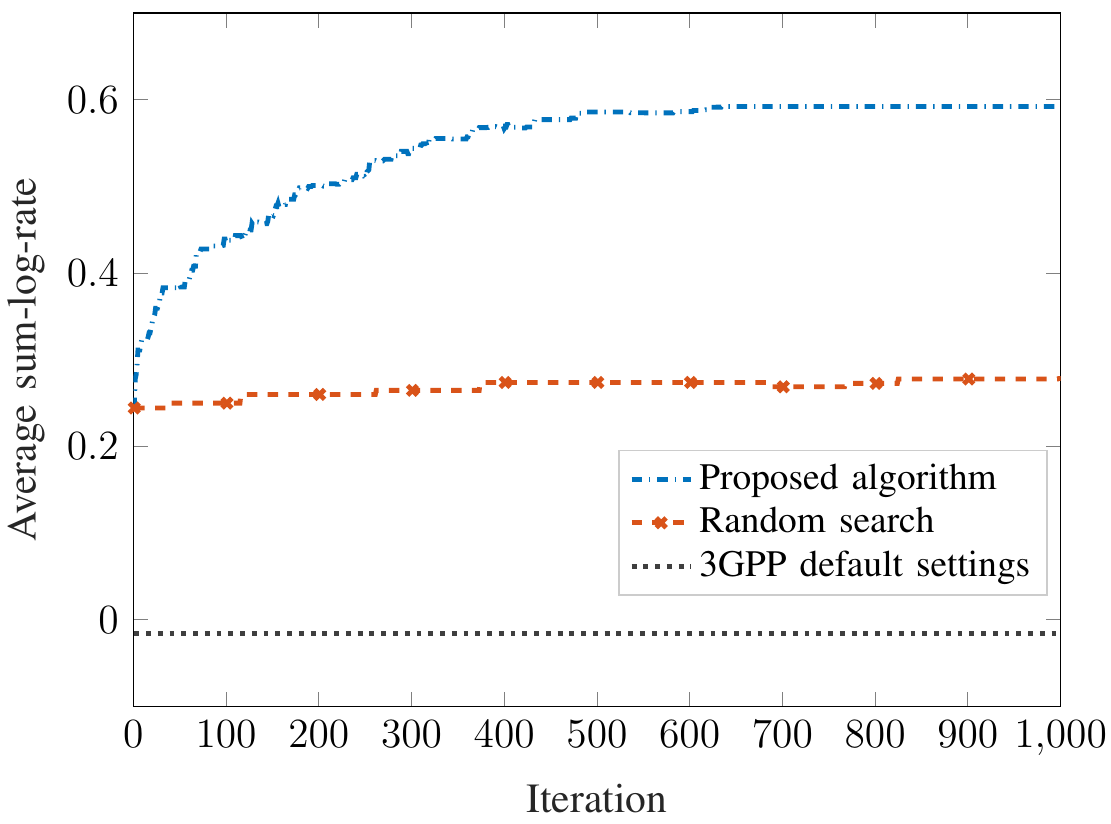}
    \subcaption{Sum-log-rate vs. iteration}
    \end{minipage}%
    \enspace
    \begin{minipage}[b]{0.5\linewidth}
    \vspace{-0.2cm}
    \centering
    \includegraphics[width=\textwidth]{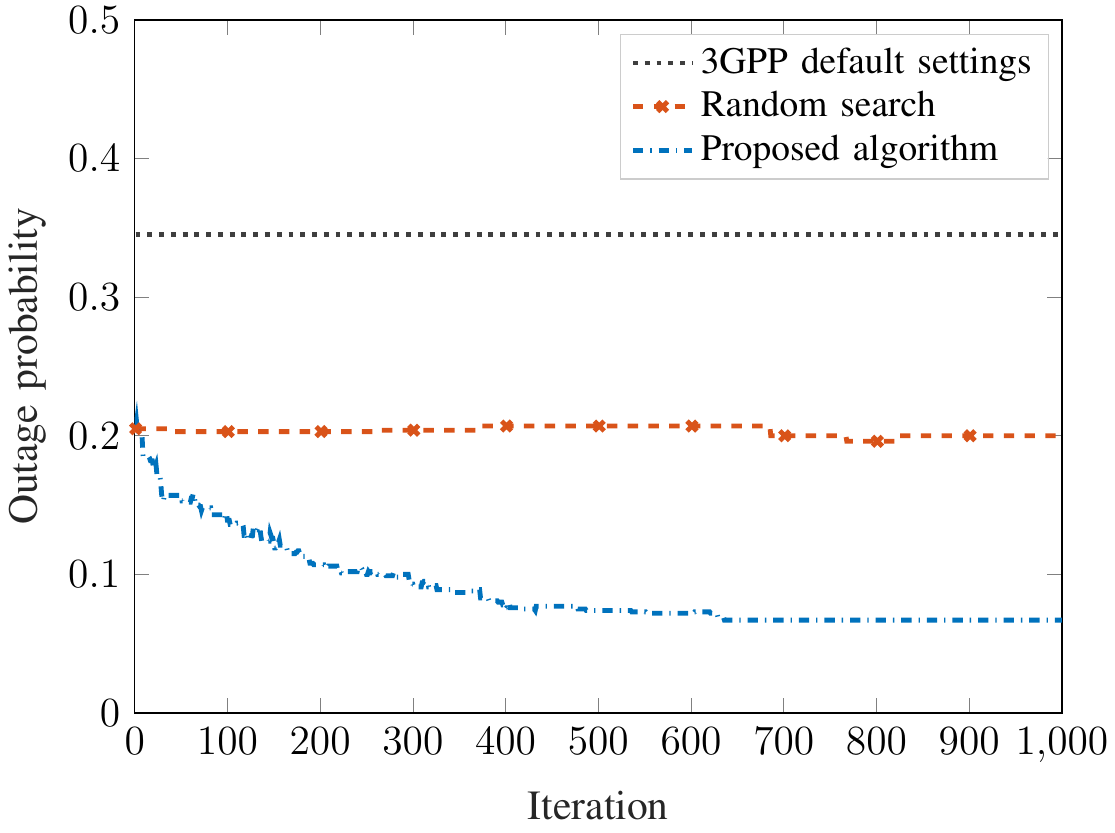}
    \subcaption{Outage probability vs. iteration}
    \end{minipage}
    \vspace{-0.2cm}
\caption{Comparison of the proposed algorithm, 3GPP default settings, random search, and Bayesian optimization in terms of average sum-log-rate and outage probability values versus iteration in Layout 2.}
\label{fig:DLonly_performance_layout2}
\vspace{-0.4cm}
\end{figure}

\textbf{Throughput and coverage gain is large compared to baselines.} Fig. \ref{fig:DLonly_performance_layout1} and \ref{fig:DLonly_performance_layout2} highlight the improvement achieved by the proposed algorithm in terms of both sum-log-rate and outage probability. It can be observed from Fig. \ref{fig:R} that the proposed algorithm reaches the peak value around $400$-th iteration and then converges while BO converges to a sub-optimal point around that iteration. Meanwhile, random search increases slowly and its value remain far below the proposed algorithm even at the $800$-th iteration. Similar observations can be made for Fig. \ref{fig:zeta}. All of the algorithms optimize the sum-log-rate and outage compared to the 3GPP default settings. However, by increasing log-utility by over 4 times and decreasing outage probability by around $90\%$, the proposed algorithm outperforms conventional BO and random search, which achieve around 2 times more log-utility and decrease outage probability by around $50\%$ and $44\%$, respectively, compared to the 3GPP default settings.

\begin{figure}[h]
    \hspace{-0.5cm}
    \begin{minipage}[b]{0.5\linewidth}
    \centering
    \includegraphics[width=\textwidth]{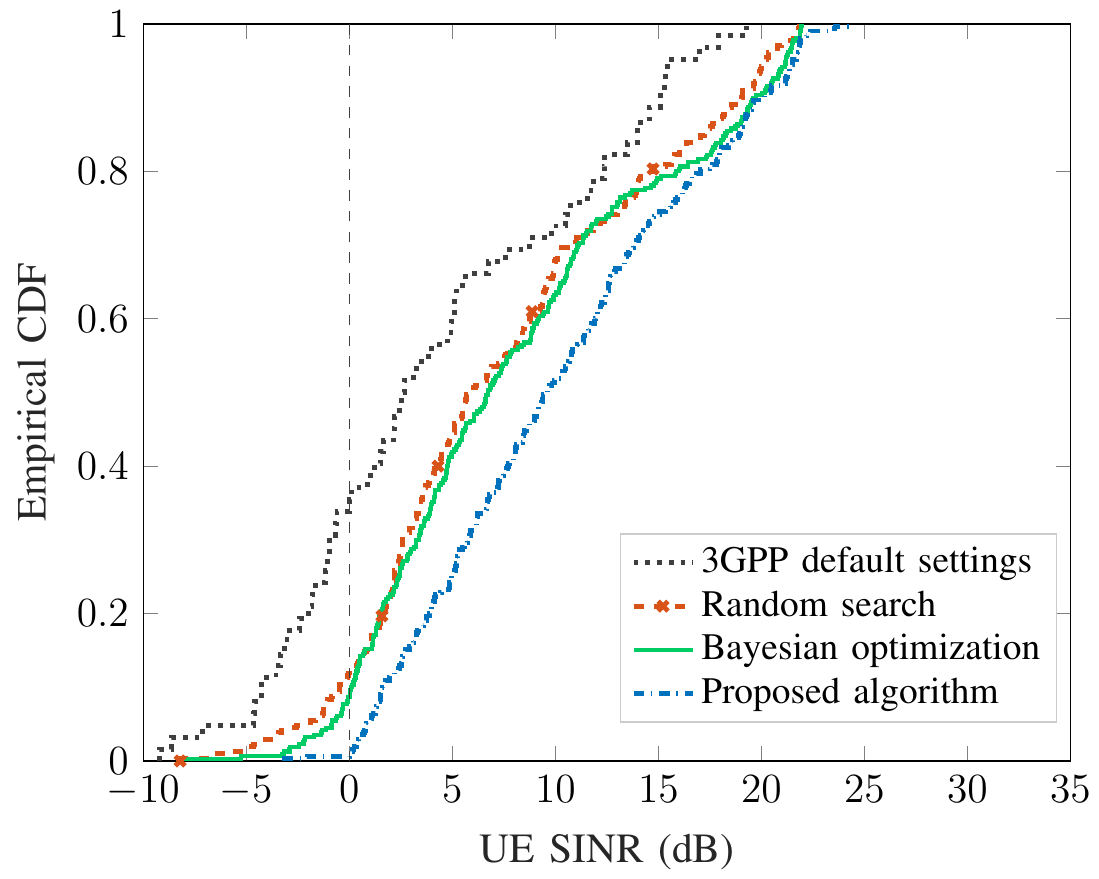}
        \subcaption{Layout 1}
        \label{fig:DLonly_UESINRa}
    \end{minipage}%
    \enspace
    \begin{minipage}[b]{0.5\linewidth}
        \centering
        \includegraphics[width=\textwidth]{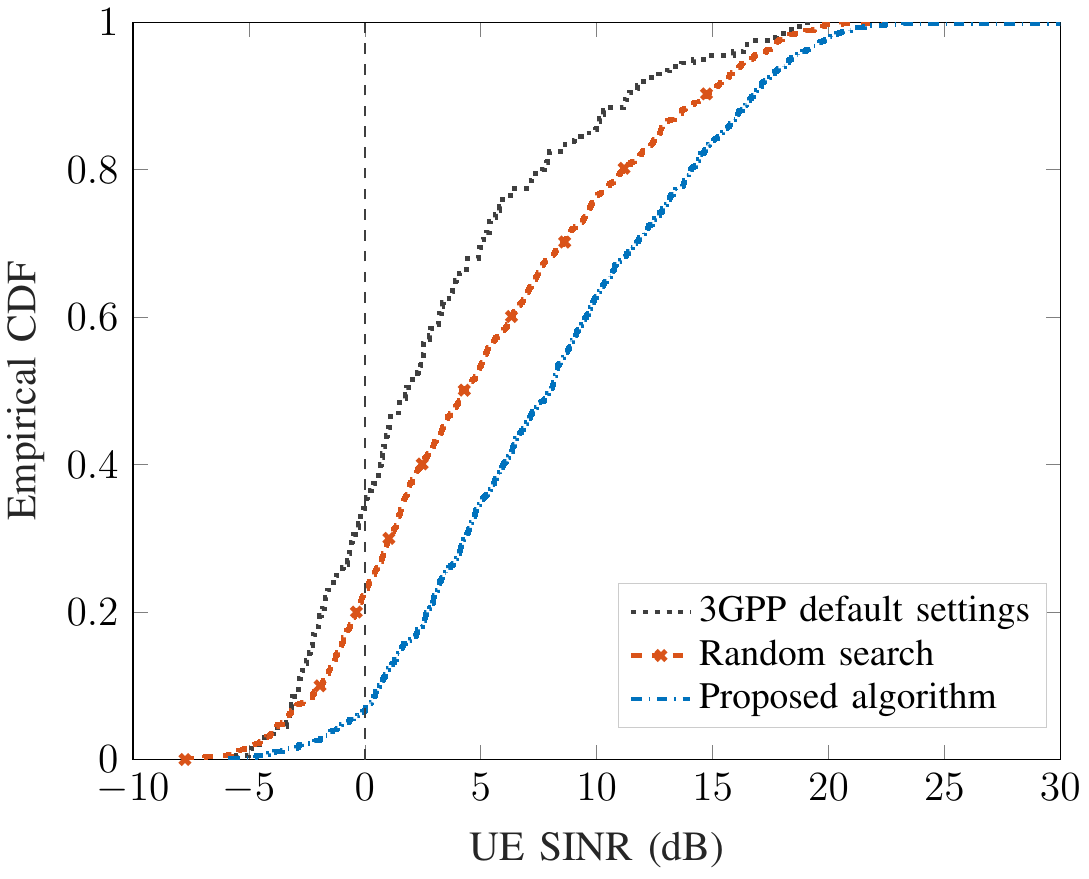}
        \subcaption{Layout 2}
        \label{fig:Layout2_UESINR}
    \end{minipage}
    \caption{Comparison of the proposed algorithm, 3GPP default settings, random search, and Bayesian optimization in terms of the empirical CDF of downlink UE SINR in Layout 1 and Layout 2.}
    \label{fig:DLonly_UESINR}
\end{figure}

\textbf{Algorithm makes a trade-off between sum-rate maximization and fairness.} The right shift in empirical CDFs of the UE SINR, Fig. \ref{fig:DLonly_UESINR}, shows that the optimal angle values found by the algorithm provide a better overall SINR distribution for UEs than the 3GPP and other comparative baselines. The effect of utilizing the sum-log-rate can be noticed from this figure that the performance improvement is not limited to a specific group of users; instead, there is an improvement across the whole range of performance metrics. This implies that the throughput and coverage gains shown in Fig. \ref{fig:DLonly_performance_layout1} and \ref{fig:DLonly_performance_layout2} are achieved by making UEs in all SINR ranges better off, highlighting the fairness aspect of the algorithm. 

\begin{figure}[h]
       \begin{minipage}[t]{.5\linewidth}
       \vspace{-0.2cm}
       \centering
       \includegraphics[width=\textwidth]{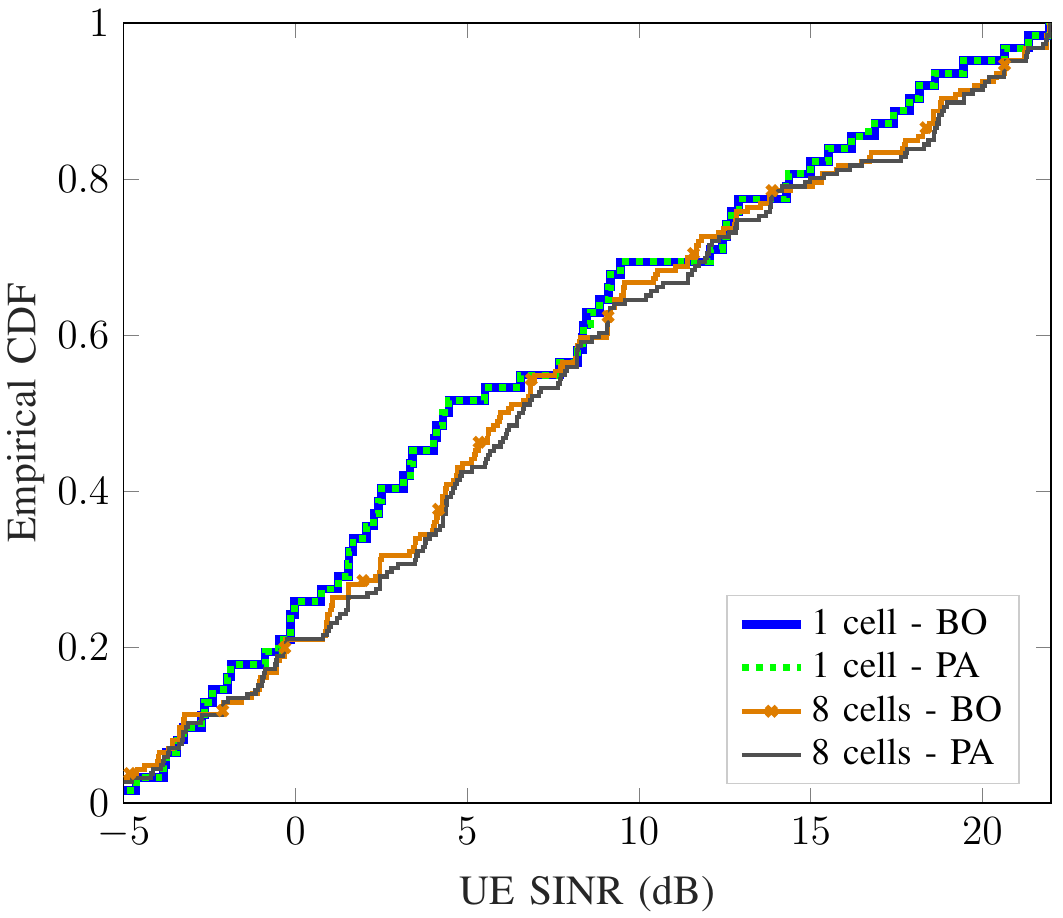}
       \subcaption{1 and 8 cells are optimized.}
       \label{fig:LessCell_UESINRa}
       \end{minipage}%
       \begin{minipage}[t]{.5\linewidth}
       \vspace{-0.2cm}
       \centering
       \includegraphics[width=\textwidth]{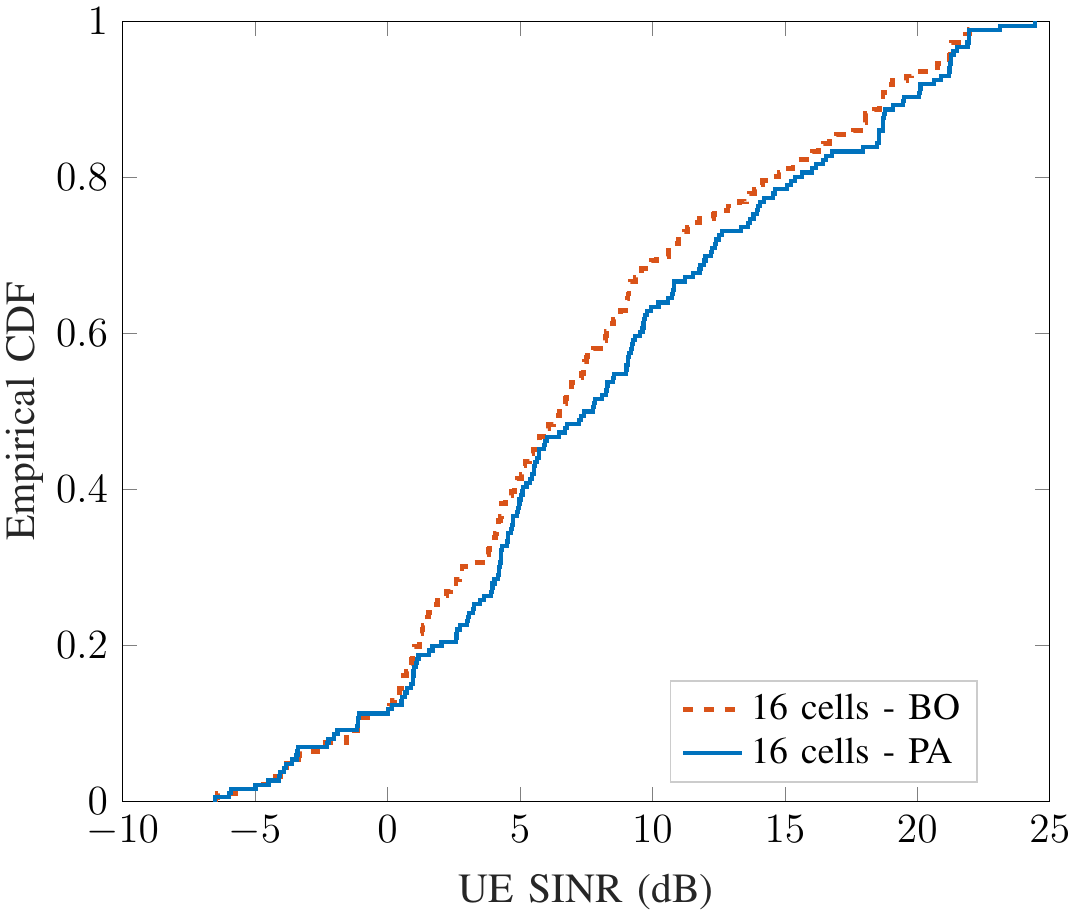}
       \subcaption{16 cells are optimized.}
       \label{fig:LessCell_UESINRb}
       \end{minipage}%
       \vspace{-0.2cm}
       \caption{Comparison of the proposed algorithm (PA) and Bayesian optimization (BO) in terms of downlink UE SINR in Layout 1. A fraction of cells (i.e., 1, 8, and 16 cells) among $M = 32$ are optimized.}
       \label{fig:LessCell_UESINR}
       \vspace{-0.4cm}
   \end{figure}

Table \ref{table:DL_UESINR} depicts the downlink UE SINR improvement that the proposed algorithm delivers in both layouts. We can observe that in Layout 1, the proposed framework has around $5.95$ dB, $2.16$ dB, and $1.55$ dB SINR improvement for the UEs in the outage range ($10\%$ SINR), and $7.12$ dB, $3.78$ dB, and $2.90$ dB increase for the median UE SINR compared to the 3GPP default settings, random search, and conventional BO, respectively. In Layout 2, these improvements are around $3.50$ dB and $2.57$ dB for the outage SINR; and $6.15$ dB, and $3.69$ dB for the median SINR compared to the 3GPP default settings and random search, respectively. BO is not used as a comparison method in this layout since the dimension of the input parameters (i.e., antenna parameters) is 231 (77 cells with 3 parameters), and conventional BO is not a suitable method for the problems of this dimension. The performance of the BO being close to random search in Layout 1 supports our claim that it is not a suitable algorithm for optimization problems with high input dimensions, as $96$ input parameters are optimized in this layout. The degradation of the BO performance with increasing dimensionality can also be observed in Fig. \ref{fig:LessCell_UESINR}. 
\begin{table}[h]
\caption{Downlink median and $10\%$ outage SINR for different optimization algorithms and layouts.}
\label{table:DL_UESINR}
\begin{minipage}[b]{0.49\textwidth}
\vspace{-0.2cm}
\centering
\begin{tabular}{m{2.8cm}>{\centering\arraybackslash}m{1.8cm}>{\centering\arraybackslash}m{1.5cm}} \toprule
    \textbf{Algorithm} & {$\boldsymbol{10\%}$ \textbf{Outage SINR (dB)}} & 
    {\textbf{Median SINR (dB)}} \\ \midrule
      \rowcolor{gray!10}
      \text{Proposed algorithm}  & 1.700 & 9.635  \\ 
      \text{Bayesian optimization} & 0.1517 & 6.737 \\
      \rowcolor{gray!10}
      \text{Random search}  & -0.4635 & 5.860 \\ 
      \text{3GPP default settings} & -4.250 & 2.517 \\ \bottomrule
\end{tabular} 
\subcaption{Layout 1}
\end{minipage}%
\quad 
\begin{minipage}[b]{0.49\textwidth}
\vspace{-0.2cm}
\centering
\begin{tabular}{m{2.8cm}>{\centering\arraybackslash}m{1.8cm}>{\centering\arraybackslash}m{1.5cm}} \toprule
    \textbf{Algorithm} & {$\boldsymbol{10\%}$ \textbf{Outage SINR (dB)}} & 
    {\textbf{Median SINR (dB)}} \\ \midrule
      \rowcolor{gray!10}
      \text{Proposed algorithm} & 0.6366 & 7.926  \\ 
      \text{Random search}  & -1.937 & 4.240 \\ 
      \rowcolor{gray!10}
      \text{3GPP default settings} & -2.860 & 1.780 \\ \bottomrule
\end{tabular}
\subcaption{Layout 2}
\end{minipage}
\end{table}

\textbf{Algorithm gracefully scales to large network sizes.} The performance results of Layout 2 with 19 macrocells with three sector antennas and 20 small cells, Fig. \ref{fig:DLonly_performance_layout2} and \ref{fig:Layout2_UESINR}, show how our algorithm successfully scales to larger network sizes while maintaining its significant gain compared to random search and 3GPP default settings. To observe the performance of the proposed algorithm with increasing network size compared to conventional BO, we also run experiments with different numbers of cells (i.e., input parameters). Fig. \ref{fig:LessCell_UESINR} illustrates the comparison between conventional BO and our proposed algorithm for the experiments that 1, 8, and 16 cells among 32 are optimized in Layout 1. To make the figure more readable, we plotted the empirical CDF of UE SINR for the cases where 16 cells are optimized in a different figure than 1 cell and 8 cells are optimized. It can be observed from the figures that the results of BO and the proposed algorithm match for the case where only 1 cell is optimized. Furthermore, in the case where 8 cells are optimized, the performance of BO and the proposed algorithm is still similar. However, as the number of optimized cells increases, our proposed algorithm starts to outperform BO. For example, in the case that 16 cells are optimized (i.e., a total of 48 input parameters are optimized in the defined problem), the proposed algorithm has a UE SINR distribution shifted to the right compared to BO and the difference is noticeable. This difference between the plots becomes significant when we optimized all the BSs in the network (i.e. 32 cells) as shown in Fig. \ref{fig:DLonly_UESINRa} earlier. We can thus infer from the figures that while our proposed algorithm continues to improve UE SINR as the number of optimized cells increases, conventional BO cannot keep track of it. Hence, the difference between the two CDF plots and average SINR values increases, showing that our proposed algorithm has better scalability to large networks.
   
\begin{figure}[h]
       \begin{minipage}[b]{0.3\linewidth}
       \centering
       \includegraphics[width=\textwidth]{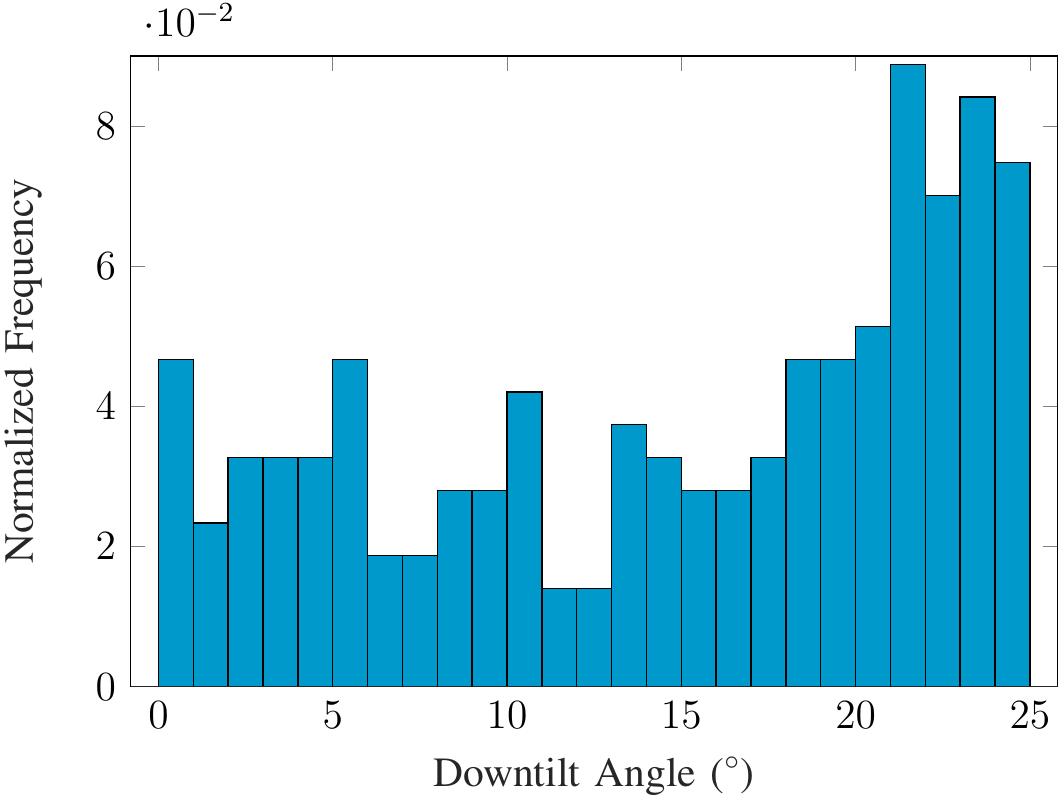}
       \label{fig:Parameters_Histogram_f}
       \end{minipage}%
       \quad \enspace
       \begin{minipage}[b]{0.3\linewidth}
       \centering
       \includegraphics[width=\textwidth]{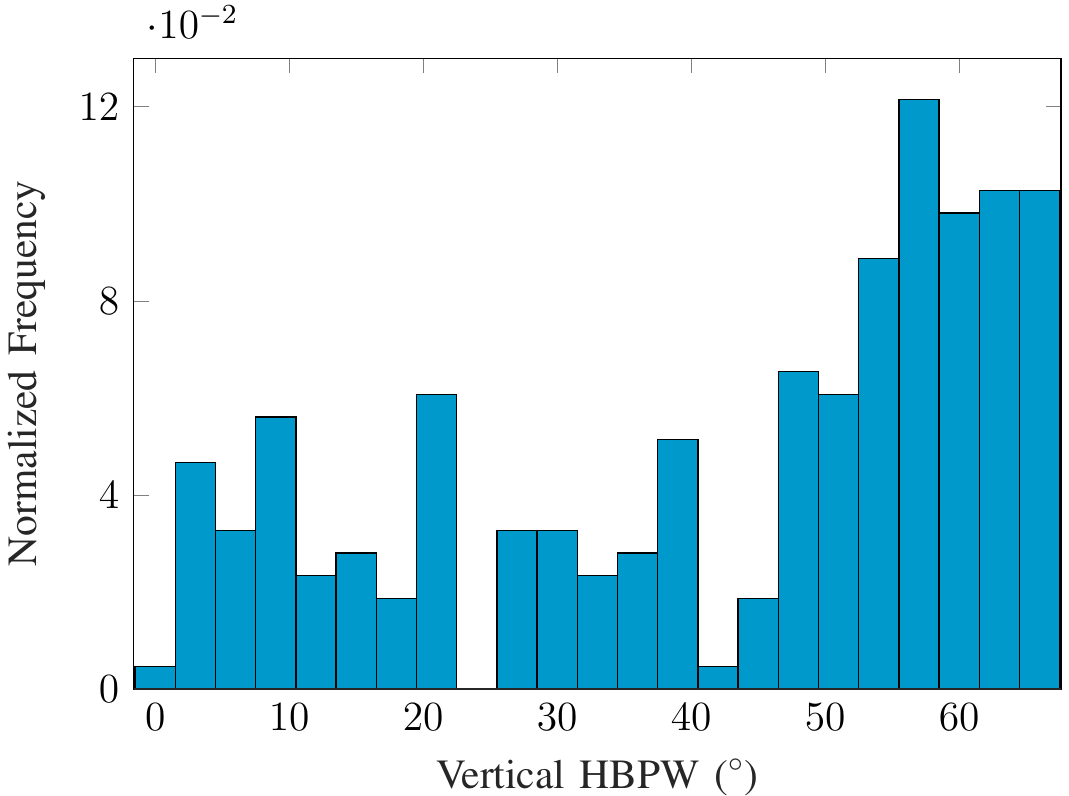}
       \subcaption{Macrocell} 
       \label{fig:Parameters_Histogram_a}
       \end{minipage}%
       \quad \enspace
       \begin{minipage}[b]{0.3\linewidth}
       \centering
       \includegraphics[width=\textwidth]{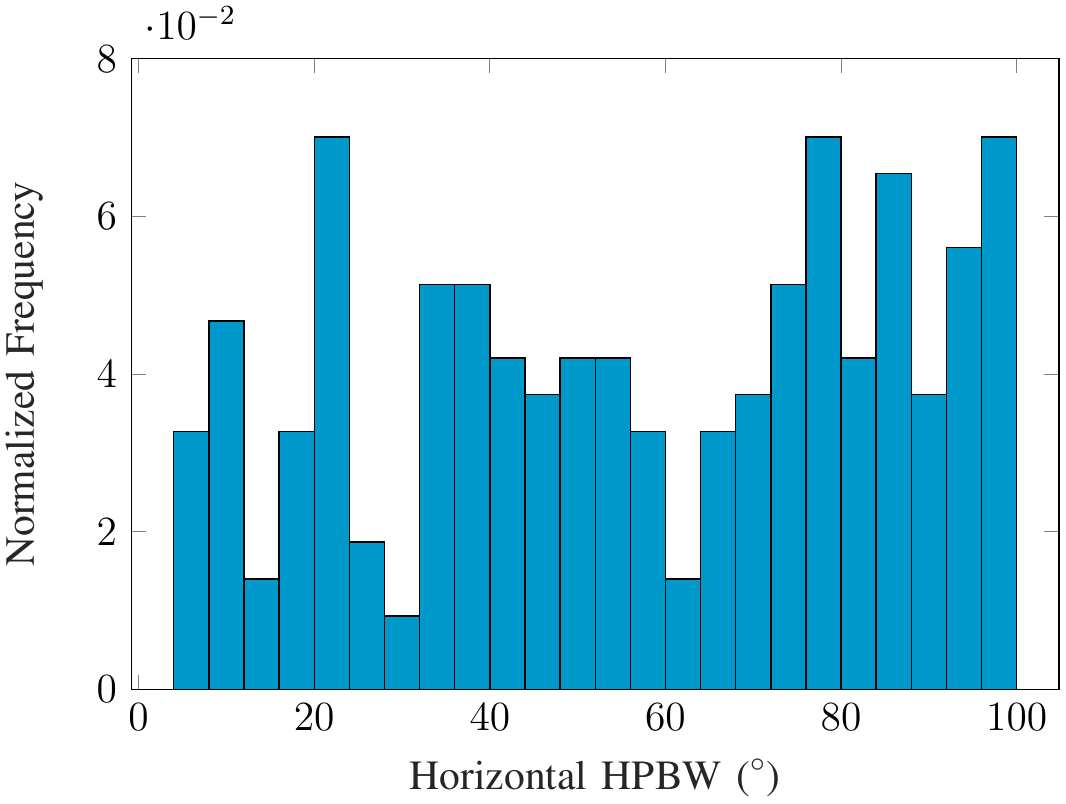}
       \label{fig:Parameters_Histogram_c}
       \end{minipage}
      \newline
      \hspace{-0.5cm}
      \begin{minipage}[b]{0.3\linewidth}
      \centering
      \includegraphics[width=\textwidth]{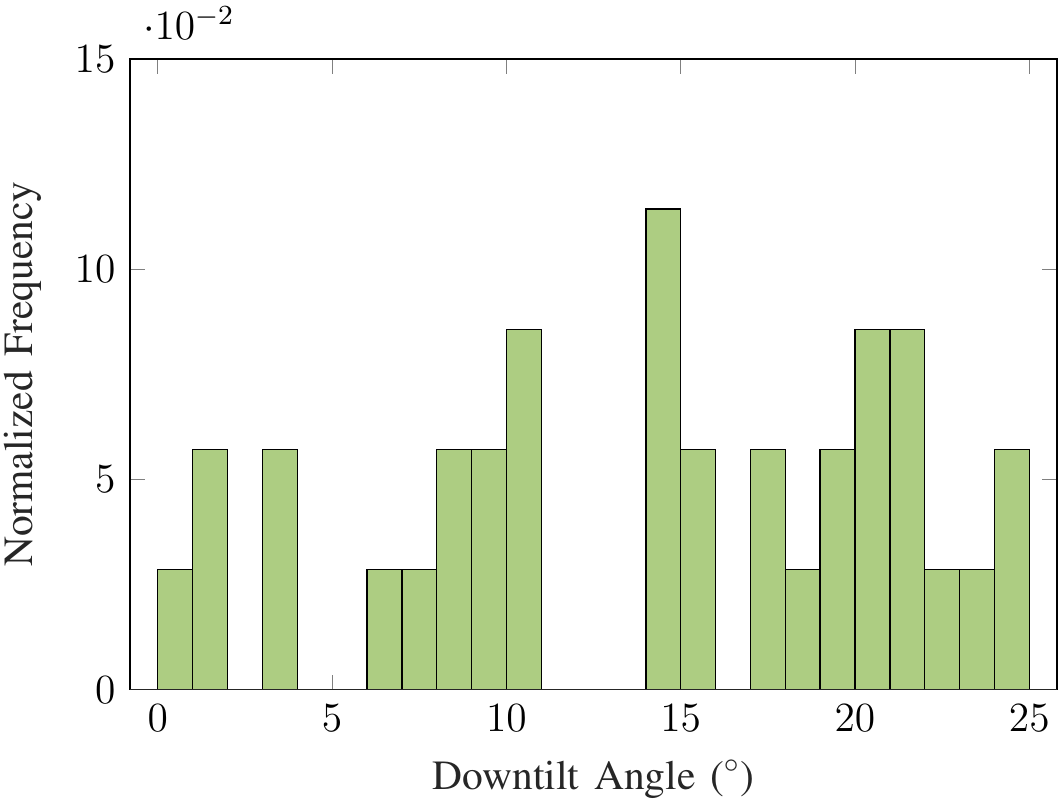}
      \label{fig:Parameters_Histogram_d}
      \end{minipage}%
      \quad \enspace
      \begin{minipage}[b]{0.3\linewidth}
      \centering
      \includegraphics[width=\textwidth]{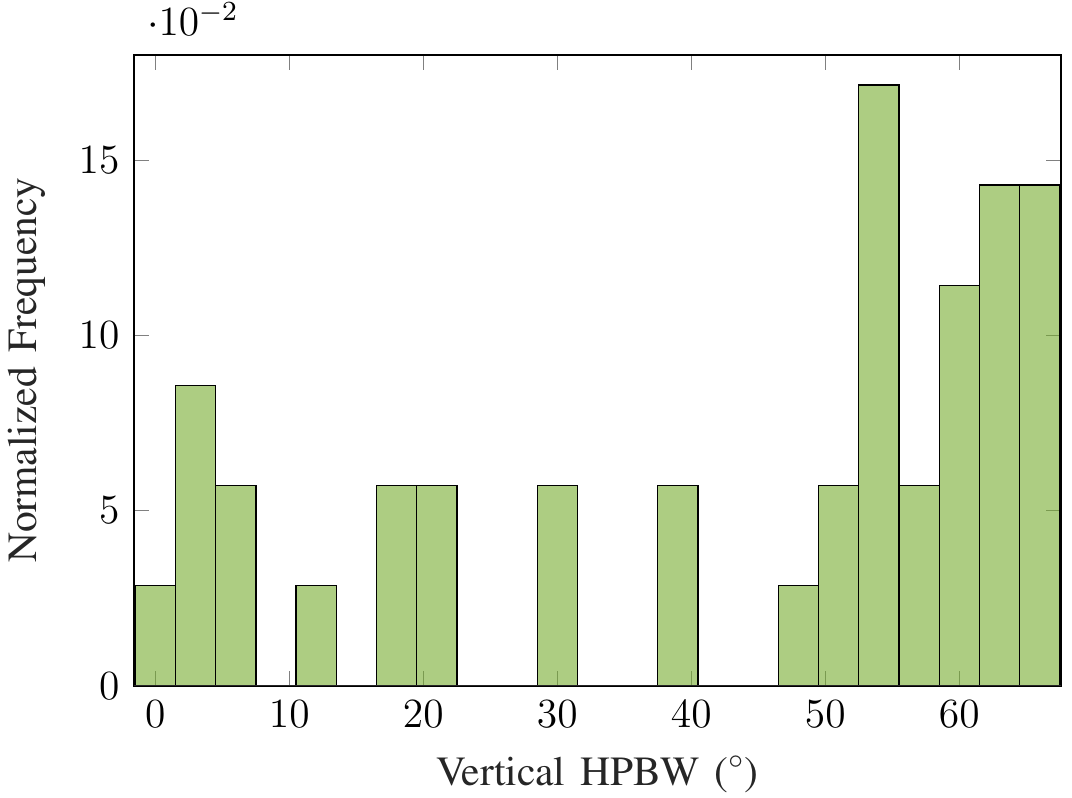}
      \subcaption{Small cell}
      \label{fig:Parameters_Histogram_b}
      \end{minipage}%
      \quad \enspace
      \begin{minipage}[b]{0.3\linewidth}
      \centering
      \includegraphics[width=\textwidth]{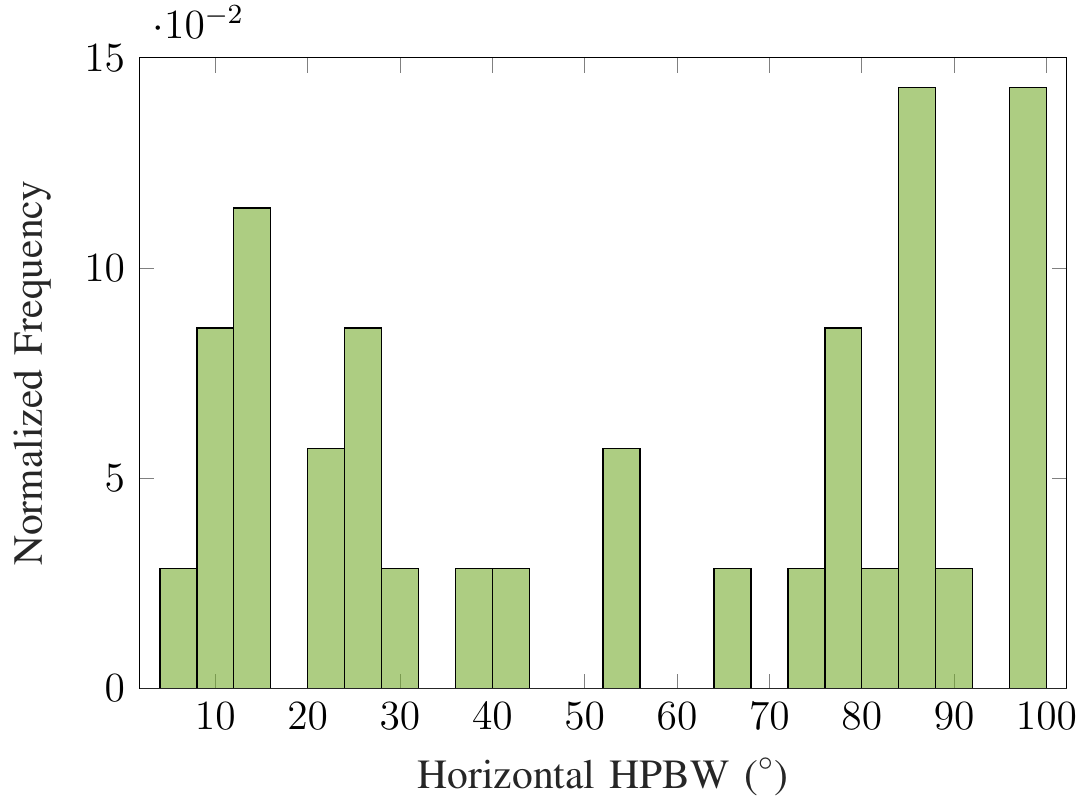}
      \label{fig:Parameters_Histogram_g}
      \end{minipage}   
       \caption{The normalized histogram of the optimal parameters found for macrocells and small cells in Layout 2.}
       \label{fig:Parameters_Histogram}
       \vspace{-0.5cm}
   \end{figure}

   Fig. \ref{fig:Parameters_Histogram} shows the histogram of optimal parameters found by the proposed algorithm over 5 realizations of Layout 2. Fig. \ref{fig:Parameters_Histogram_a} and \ref{fig:Parameters_Histogram_b} present the macrocell and small cell parameters, respectively. Only cells that are associated with a UE are included in these histograms. For cells not associated with any UE, we observe that most of them avoid interfering with other cells by choosing small vertical and/or horizontal HPBWs. For the other cells, we can observe from the histograms that antenna parameters of both macrocells and small cells mainly concentrate on the higher values. This tendency is related to the fact that we have a dense network with a large number of UEs. So, most of the cells are serving multiple UEs, and they are trying to cover all of their users by creating large beams. For example, the cell located around $(735,352)$ with azimuth $240^{\circ}$ sets its vertical HPBW to $59.5^{\circ}$ and its horizontal HPBW to $99.7^{\circ}$. The width of the beam becomes narrower for the cells with less number of UEs. They take advantage of focusing the energy on the served users and having a narrower beam, avoiding interference with neighboring cells. For example, the cell located around $(735,352)$ with azimuth $0^{\circ}$ sets its vertical HPBW to $8.82^{\circ}$ and its horizontal HPBW to $11.0^{\circ}$. 
   
   Besides the number of associated UEs and antenna azimuth, BS height also affects the antenna beam pattern. For example, since the height of the small cells is smaller and closer to the height of the UE, it is expected that they have smaller downtilt values and larger HPBW on the vertical plane compared to macrocells in order to serve more UEs.
  
\subsection{Trade-off Between Capacity and Coverage}
\begin{figure}[ht]
   \centering
   \includegraphics[width=0.6\textwidth]{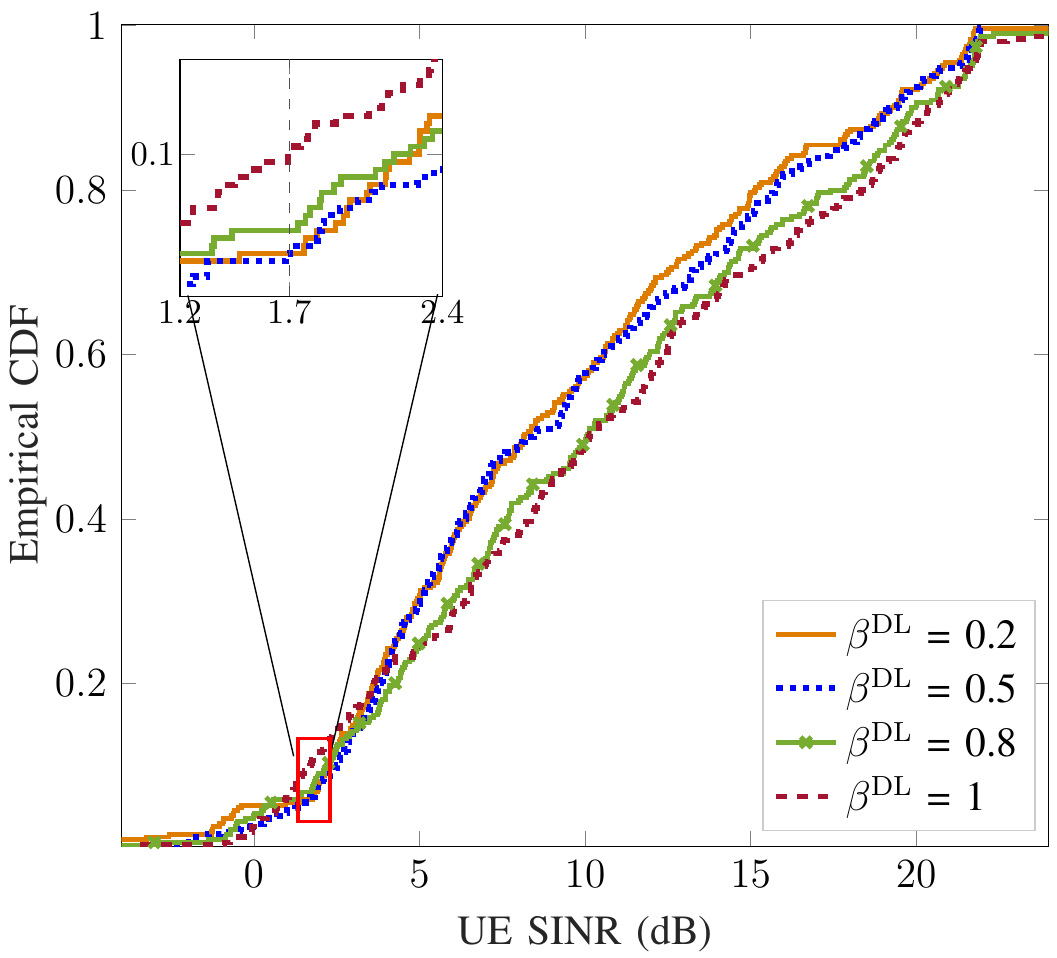}
   \vspace{-0.2cm}
\caption{Empirical CDF plot of downlink UE SINR evaluated with different $\beta^{\rm{DL}}$ values on Layout 2, $\mathsf{T} = 1.7$ dB.}
\label{fig:DL_differentbeta}
\vspace{-0.5cm}
\end{figure}

To observe how two objectives (i.e., coverage and capacity) compare, we experiment with the $\beta^{\rm{DL}}$ coefficient after setting $\alpha = 0$. Fig. \ref{fig:DL_differentbeta} shows the empirical CDF of UE SINR for $\beta^{\rm{DL}}$ values of $0.2$, $0.5$, $0.8$, and $1$. In this experiment, we use $\beta^{\rm{DL}} = 1$ plot to determine the $10\%$ SINR and set the $\mathsf{T}$ threshold to this  value in order to better observe the effect of smaller $\beta^{\rm{DL}}$ on the outage optimization (i.e., lower SINR values). In the case of $\beta^{\rm{DL}} = 1$, we are trying to optimize the sum-log-rate of users without considering how many users are in the outage. Hence, the algorithm can sacrifice some users to increase the overall sum-log-rate. When $\beta^{\rm{DL}}$ is set to small values such as $0.2$, the optimization problem is mostly solved for the outage. Thus, the main objective of the algorithm is to make outage probability as close to $0$ as possible, and it is nearly indifferent to SINR increase if it is already higher than the threshold. Fig. \ref{fig:DL_differentbeta} confirms this as the $\beta^{\rm{DL}} = 0.2$ plot is the leftmost while its outage probability is the lowest compared to the other $\beta^{\rm{DL}}$ values. On the other hand, $\beta^{\rm{DL}} = 1$ results in the best SINR performance while its outage probability is the highest. For the other $\beta^{\rm{DL}}$ values, UE SINR CDF plot is in-between $\beta^{\rm{DL}} = 0.2$ and $\beta^{\rm{DL}} = 1$. We can conclude from the figure that for the given layout, $\beta^{\rm{DL}} = 0.8$ makes an effective trade-off between sum-log-rate and outage probability as its CDF curve is very close to the curve with $\beta^{\rm{DL}} = 1$, and also it has an outage probability that is close to the one with $\beta^{\rm{DL}} = 0$. 
\vspace{-0.2cm}
\subsection{Uplink and Downlink Joint Optimization Performance}
\begin{figure*}[h]
         \hspace{-0.6cm}
         \begin{minipage}[b]{.5\linewidth}
         \centering
         \includegraphics[width=\textwidth]{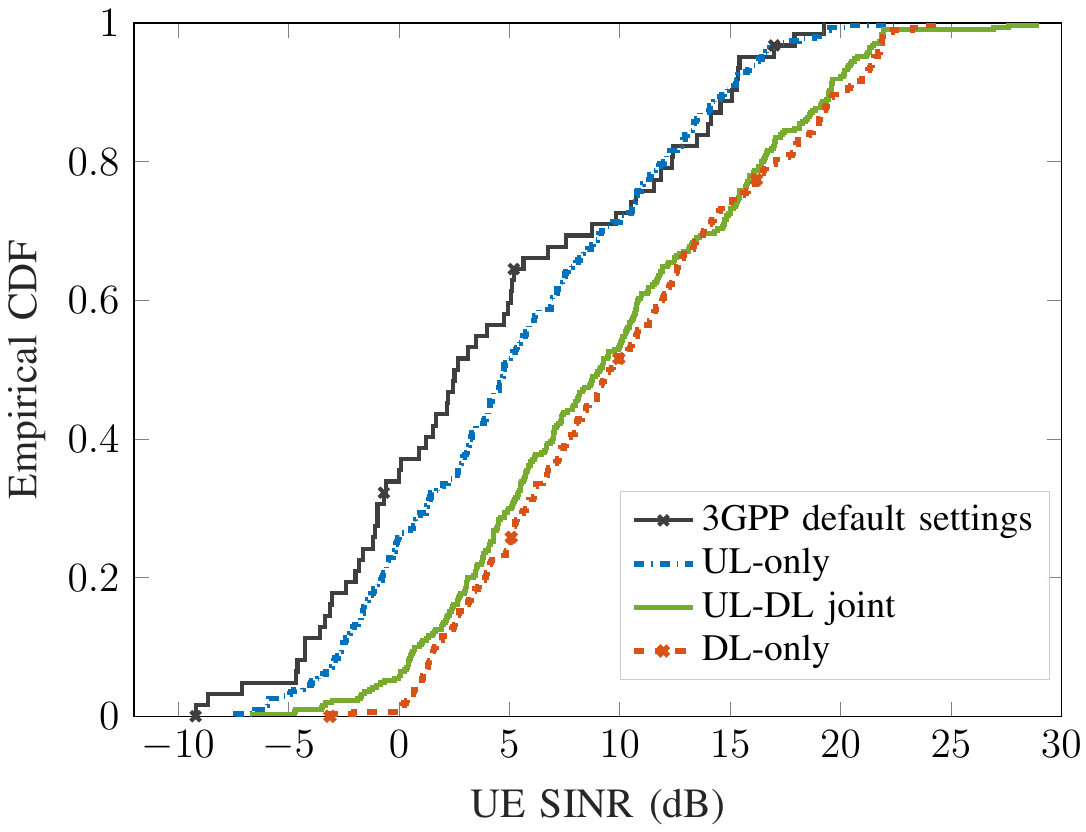}
         \subcaption{Downlink UE SINR}
         \label{fig:alpha05_UESINR_DL}
         \end{minipage}%
         \enspace
         \begin{minipage}[b]{.5\linewidth}
         \centering
         \includegraphics[width=\textwidth]{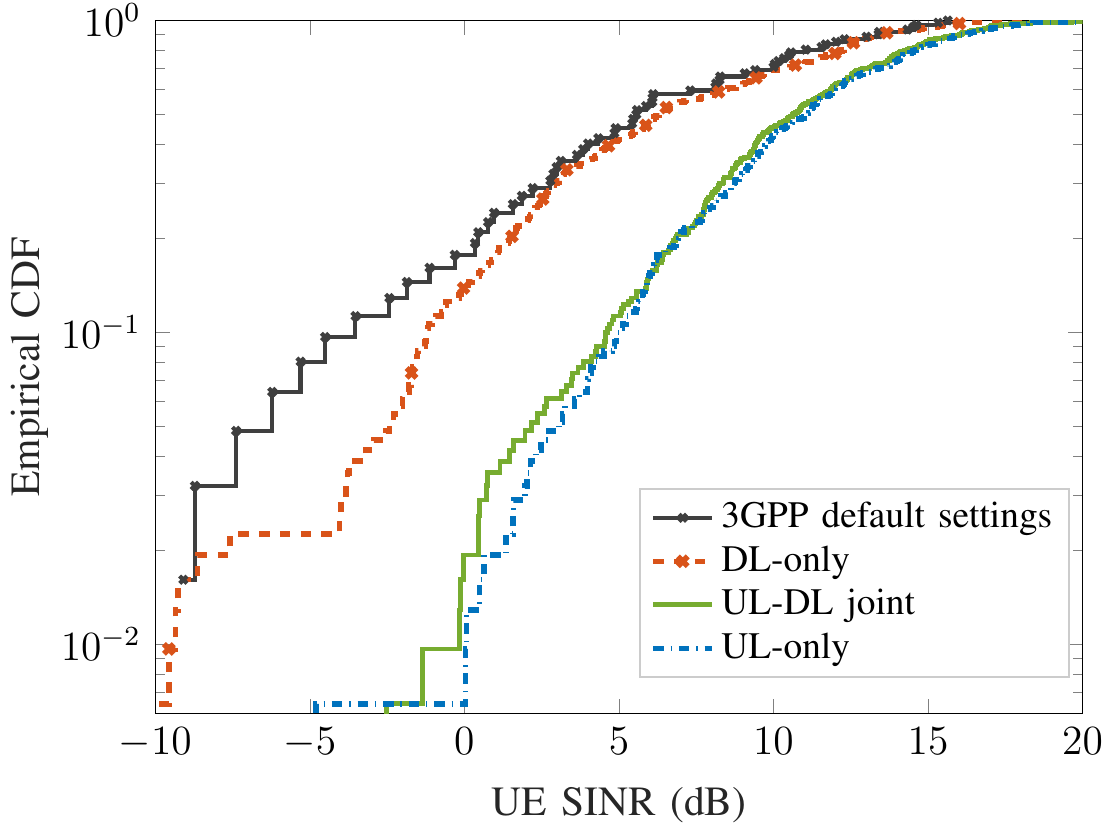}
         \subcaption{Uplink UE SINR}
         \label{fig:alpha05_UESINR_UL}
         \end{minipage}
         \vspace{-0.2cm}
         \caption{Empirical CDF plot of UE SINR values after optimization for DL-only, UL-only, and UL-DL joint in Layout 1 for $\alpha = 0.5$.}
         \label{fig:alpha05_UESINR}
         \vspace{-0.5cm}
   \end{figure*}

\begin{figure*}[h]
       \hspace{-0.6cm}
       \begin{minipage}{.5\linewidth}
       \centering
       \includegraphics[width=\textwidth]{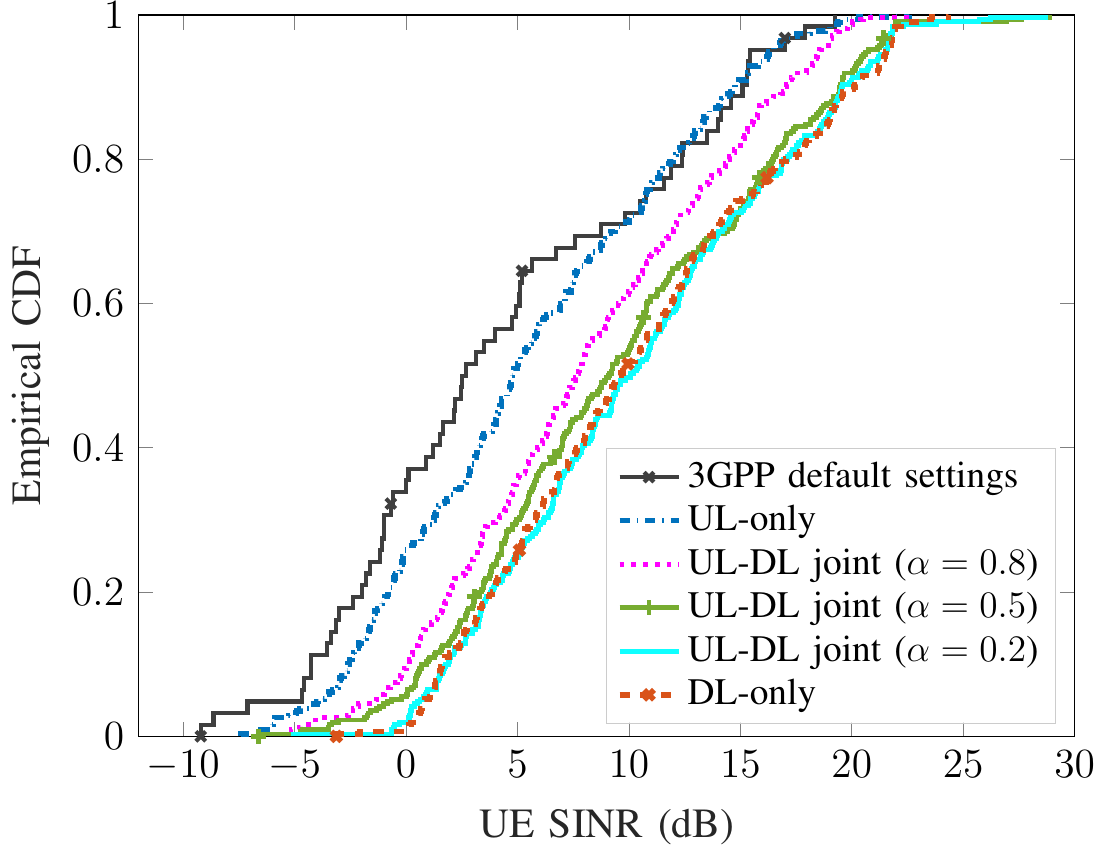}
       \subcaption{Downlink UE SINR}
       \label{fig:alpha02_UESINR_DL}
       \end{minipage}%
       \enspace
       \begin{minipage}{.5\linewidth}
       \centering
       \includegraphics[width=\textwidth]{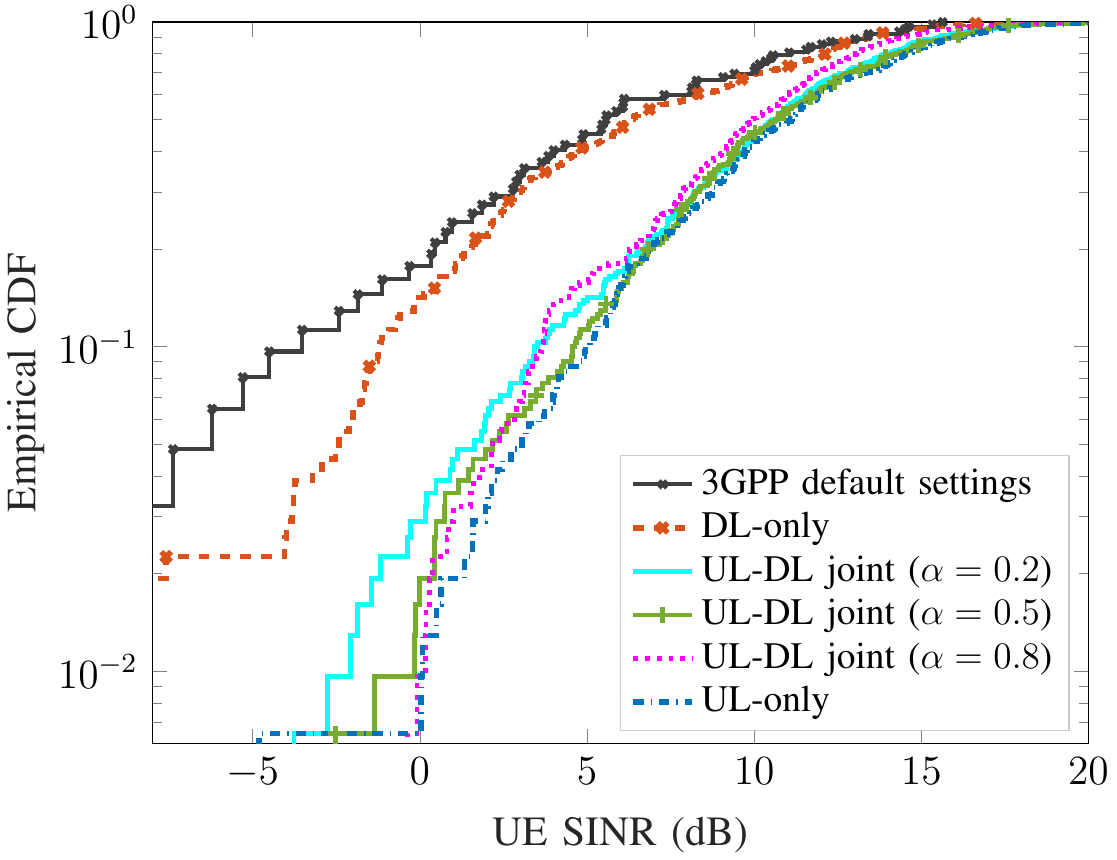}
       \subcaption{Uplink UE SINR}
       \label{fig:alpha02_UESINR_UL}
       \end{minipage}
       \caption{Empirical CDF plot of UE SINR values after optimization for DL-only, UL-only, and UL-DL joint in Layout 1 for different UL weightings $\alpha$.}
       \label{fig:alpha02_UESINR}
       \vspace{-0.8cm}
 \end{figure*}

To investigate how the parameter choice differs in uplink and downlink directions, we first perform downlink-only and uplink-only optimizations and observe whether the optimal angle and beamwidth values differ significantly and frequently on a cell-by-cell basis in these two cases. We then perform an uplink and downlink joint optimization to find the best values per cell and the loss from the optimum downlink or uplink values. Fig. \ref{fig:alpha05_UESINR} and \ref{fig:alpha02_UESINR} depict the distribution of UE SINR achieved after uplink (i.e., UL-only), downlink (i.e., DL-only), and uplink and downlink joint optimization (i.e., UL-DL joint). The uplink weight $\alpha$ is set to $0.5$ in the joint optimization shown in Fig. \ref{fig:alpha05_UESINR} while its value varies in Fig. \ref{fig:alpha02_UESINR}. In both cases, the outage threshold, $\mathsf{T}$ is set as $0$ dB, and the trade-off coefficients between rate and outage probability, $\beta^{\rm{UL}}$ and $\beta^{\rm{DL}}$, are chosen as $0.2$ and $0.8$, respectively. These $\beta$ values are chosen to focus the optimization more on data rates in downlink and coverage in uplink transmission considering the limited UE transmit
power. To better monitor these desired improvements, we present uplink UE SINR CDF on a log scale while using a linear scale for downlink UE SINR. 

\textbf{Throughput and coverage gain is large compared to single direction optimization.}
It is evident from Fig. \ref{fig:alpha05_UESINR} that there is a significant performance loss due to single direction optimization. UL-only optimization degrades the UE SINR performance of downlink transmission while uplink performance is degraded with DL-only optimization. This is because DL-only (UL-only) optimization only considers the performance of the downlink (uplink) direction to set the antenna parameter and disregards the uplink (downlink) performance. On the other hand, joint optimization increases uplink and downlink median and $10\%$ outage SINR compared to the DL-only and UL-only optimization, respectively. The quantitative results for $10\%$ outage and median SINR show that there is an increase in uplink median and $10\%$ outage SINR by around 4.28 dB and 5.82 dB, respectively, compared to the DL-only optimization while downlink median and $10\%$ outage SINR is increased by around 4.38 dB and 3.47 dB compared to the UL-only optimization in the given layout. The better improvement of uplink $10\%$ outage SINR and downlink median SINR compared to the uplink median and downlink $10\%$ outage SINR, respectively, highlights the effect of the chosen $\beta^{\rm{UL}}$ and $\beta^{\rm{DL}}$ values. These numerical results and figures indicate that UL-DL joint optimization with the chosen $\beta$ values greatly improves the uplink coverage and the downlink throughput as desired.

\vspace{-0.2cm}
\subsection{Trade-off Between Uplink and Downlink Optimization}
We experiment with $\alpha$ values in optimization problem \eqref{eq:optimization_prob_ULDL} to understand how the optimal parameter values and the loss differ compared to the case where $\alpha = 0.5$, Fig. \ref{fig:alpha05_UESINR}. The comparison plots, which include the results of both the single and joint direction optimization, are presented in Fig. \ref{fig:alpha02_UESINR} for different $\alpha$ values. Fig. \ref{fig:alpha02_UESINR_DL} depicts the distribution of downlink UE SINR while Fig. \ref{fig:alpha02_UESINR_UL} shows the uplink UE SINR distribution. In both cases, we can observe that aligning with the objective, UL-DL joint optimization curve approaches to uplink optimization curve as $\alpha$ increases, which means that joint optimization mainly optimizes uplink transmission. On the other hand, it approaches the downlink optimization curve as $\alpha$ decreases. Hence, by optimizing the $\alpha$ value according to the needs of the different networks, one can find a good compromise between uplink and downlink transmission, which will make the performance of the overall transmission better. We can infer from the figure that for the given layout, $\alpha = 0.5$ provides a good balance between uplink and downlink, which effectively optimizes for downlink capacity and uplink coverage. However, note that UL-DL joint optimization performs significantly better than the UL-only (DL-only) optimization even with $\alpha = 0.8$ ($\alpha = 0.2$), achieving better overall downlink (uplink) UE SINR.
\vspace{-0.4cm}
\subsection{Complexity and Efficiency Analysis} \label{sec:CompAndEfecAnalys}
We also compare our algorithm with an RL method -- DDPG -- and BO using the problem formulation and simulation environment of the work \cite{dreifuerst2021optimizing} which is provided at \cite{dreifuerstcode}. More specifically, in this simulation, an RSRP map which is simulated using QuaDRiGa is used by considering a network with 5 BSs with three sector antennas. Besides, instead of downtilt angle and HPBW optimization as in our earlier problem formulation, the formulation in \cite{dreifuerst2021optimizing} is followed, and downtilt angle and transmit power are optimized using their defined optimization function, which tries to minimize both under-coverage (i.e., locations that do not have enough received signal strength) and over-coverage (i.e., locations where the interfering cells are generating too much interference). The possible values for downtilt are discrete in the range $[0,10]^{\circ}$ and transmit power has continuous values in the range $[30, 50]$ dB. For more details about the simulation environment and the problem formulation, please see \cite{dreifuerst2021optimizing}. In this paper, we use these comparisons to show the sample efficiency and time efficiency of our algorithm compared to RL and BO, respectively.
\begin{figure}[h]
    \begin{minipage}[t]{0.5\linewidth}
    \centering
    \includegraphics[width=\textwidth]{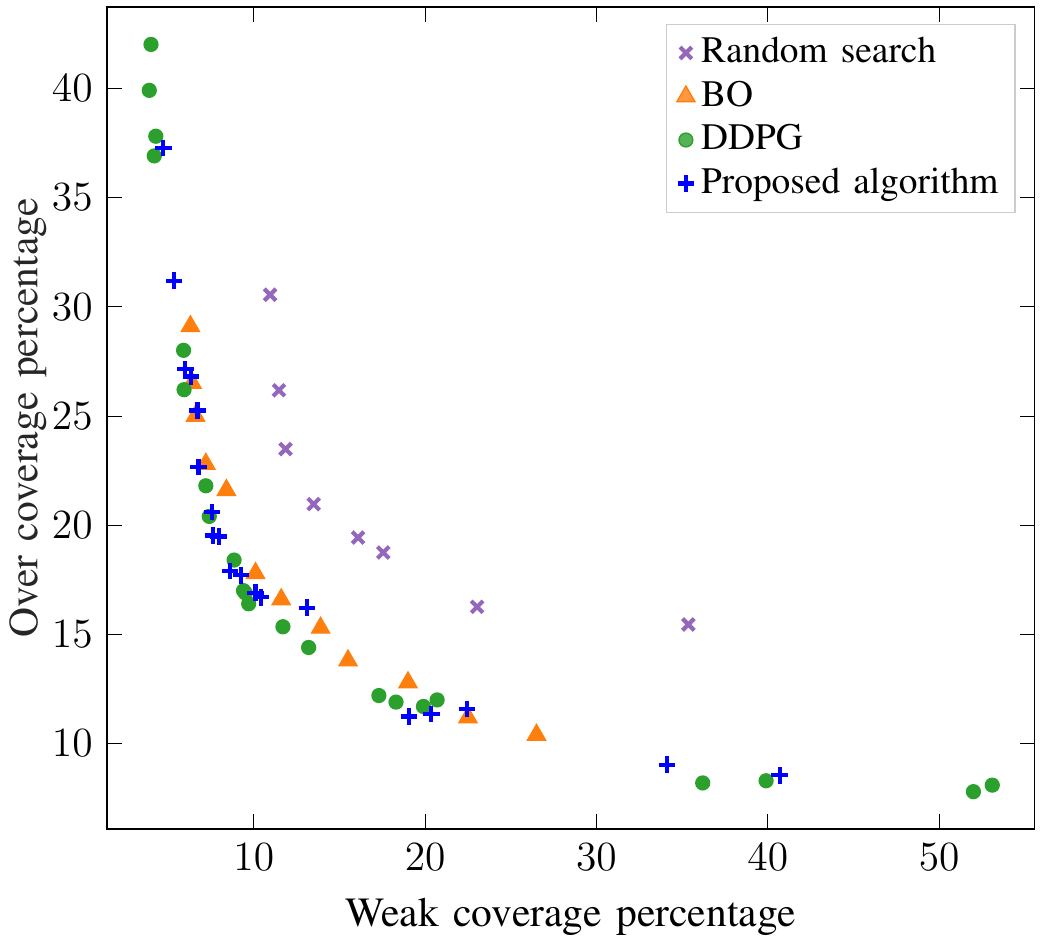}
    \vspace{-0.7cm}
    \subcaption{Sample efficiency}
    \label{fig:sample_efficiency}
    \end{minipage}
    \enspace
    \begin{minipage}[t]{0.5\linewidth}
    \centering
    \includegraphics[width=\textwidth]{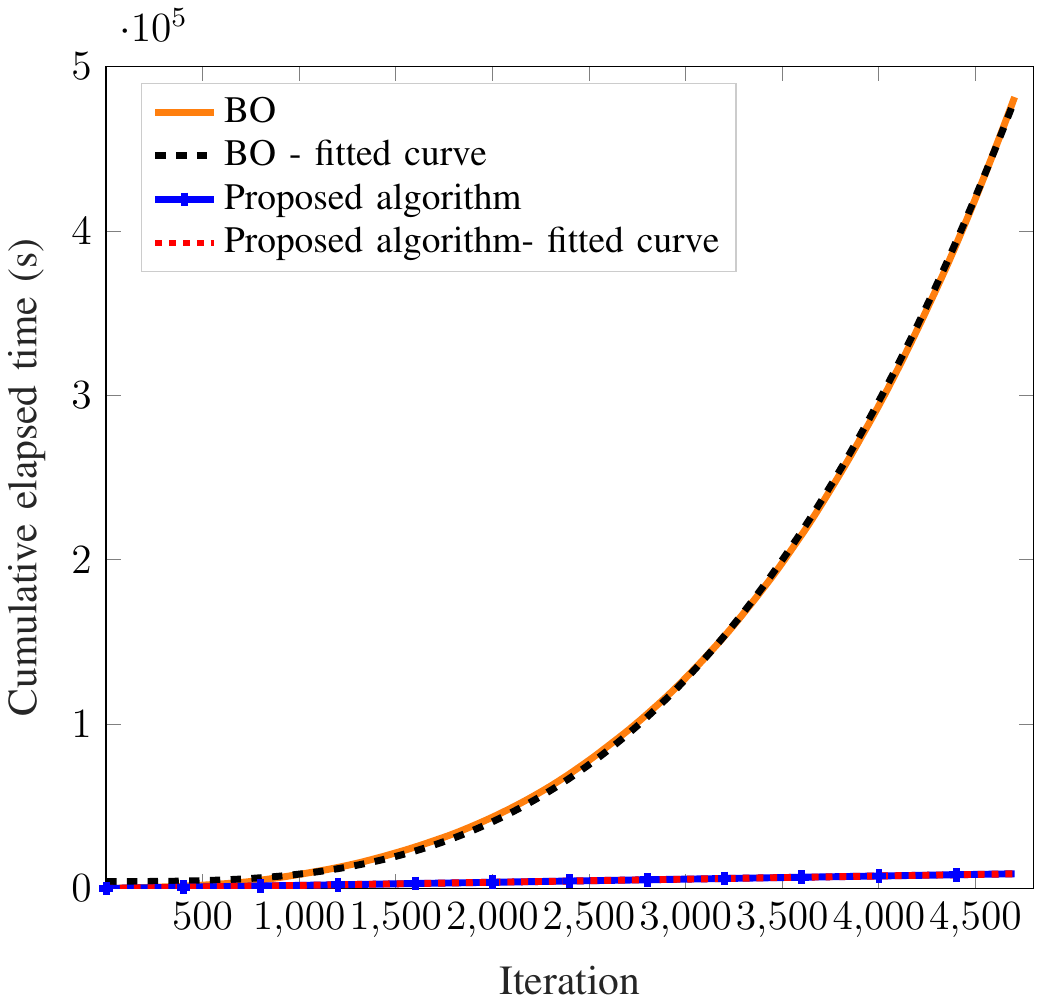}
    \subcaption{Time complexity}
    \label{fig:time_efficiency}
    \end{minipage}
    \vspace{-0.2cm}
\caption{Comparison of the proposed algorithm, BO, and DDPG in terms of sample and time efficiency.}
\label{fig:complexity_analysis}
\end{figure}

BO and DDPG frontiers in Fig. \ref{fig:sample_efficiency} are plotted using the results in \cite{dreifuerst2021optimizing}, which are obtained by using 1012 evaluations (512 for initialization and 500 for optimization) and  600,000 evaluations, respectively. We first analyzed our proposed algorithm using the same number of evaluations (for both initialization and optimization) with BO and observed that the proposed algorithm finds a better frontier compared to BO and mostly matches with the RL frontier. We then decreased the number of evaluations in initialization for our proposed algorithm to increase the time and sample efficiency. Fig. \ref{fig:sample_efficiency} shows the performance of the proposed algorithm after 750 evaluations compared to random search, BO, and DDPG. We can observe from the figure that 750 evaluations (200 for initialization and 550 for optimization) are enough to outperform BO and mostly match the performance of DDPG. These results show that both BO and our approach improve sample efficiency by over two orders of magnitude relative to DDPG, and our approach can achieve the performance of DDPG with this significant difference in the number of evaluations. 

Decreasing initialization evaluations to 200 makes a notable difference and helps increase time-efficiency further compared to BO since as it is mentioned earlier, the Gaussian process regression method, which is used in both our proposed approach and BO, has cubic computational complexity, $\mathcal{O}(\rm{T}^3)$, where $\rm{T}$ is the number of training data samples. Hence, when we use 200 evaluations for initialization, it means that $\rm{T}=200$ in our proposed algorithm throughout the optimization evaluations, and $\rm{T}$ starts from 200 and keeps increasing with each iteration in BO. This means that at every iteration, the computational time of BO is increasing, which is another dimensional restriction that BO has. This phenomenon can be clearly observed from Fig. \ref{fig:time_efficiency} which depicts the cumulative elapsed time for model building and finding a new candidate at each iteration for both BO and our proposed algorithm. In each method, 200 evaluations are used for initialization, and expected improvement acquisition function is used for BO. We show that the cumulative elapsed time of BO can be well approximated by a cubic function (i.e., $ax^{b} + c$, where $a = 4.3 \cdot 10^{-6}, b = 3.01$ and $c = 4223$), shown as a dashed line in Fig. \ref{fig:time_efficiency} while our proposed algorithm demonstrated to have linear time complexity  (i.e., $ax^{b} + c$, where $a = 1.71, b = 1.01$ and $c = 2067$). We can also infer that starting the algorithms with 512 evaluations instead of 200 increases the time elapsed at each iteration, thus shifting both curves up. This shows the importance of having good performance with less number of evaluations.

Note that in these simulations, the dimension of the optimization problem is low (i.e., 5 BSs with three sector antennas and 30 input parameters to optimize) and since we are using the exact same problem definition of the work \cite{dreifuerst2021optimizing}, we do not create a different model for each UE, and thus we cannot take advantage of that. It is thus expected that there are not much performance difference between the algorithms. We have already shown how successfully our algorithm scales to larger networks and those cases are the ones we expect more performance gains compared to the other approaches, but this experiment is also important in terms of showing the sample and time efficiency of our algorithm compared to the other baselines.
\vspace{-0.2cm}
%
%
\section{Conclusion and Future Work} \label{sec:Conc_FW}
In this paper, we study the uplink and downlink joint capacity and coverage optimization (CCO) by tuning parameters -- downtilt angle, vertical half-power beamwidth (HPBW), and horizontal HPBW -- of each cell's antenna array. We provide a data-driven framework that is suitable for real-world implementation for this non-convex and challenging problem. Considering the importance of sample and time efficiency in real-world antenna optimization, we  show that our proposed algorithm is significantly sample-efficient compared to the RL approach (i.e.,  deep deterministic policy
gradient), and time-efficient compared to conventional Bayesian optimization (BO). Using the state-of-the-art system-level simulator developed by AT\&T Labs and the layout with 19 macrocells (whose locations are based on real-world deployment) and 20 small cells, we also show that our algorithm successfully scales to large network sizes while preserving its gain compared to comparative baselines -- 3GPP default settings, random search, and conventional BO. Moreover, numerical results from the experiments on the simulator indicate that the downlink throughput and uplink coverage can be greatly improved by performing joint optimization compared to the single direction optimization. Overall, this paper demonstrates that CCO problems that only optimize the downlink have poor uplink performance, and there are significant gains to be harvested from site-specific data-driven antenna parameter optimization which can be achieved in a fast, scalable, and automated fashion. 

Future work could analyze the robustness of the proposed method over time to different UE locations and mobility, and it could include load-aware user associations. Extensions to higher carrier frequencies, larger antenna arrays, and other transmission modes would also be of interest. Moreover, our proposed algorithm can be combined with more advanced different initialization approaches to further increase sample efficiency by reducing its convergence time. Finally, given that we have provided a practically applicable learning framework in this paper, an important extension is to consider how to take this framework into a real cellular system (e.g., when and how often to train the algorithm and change parameter settings).

\section*{Acknowledgment}
The authors would like to thank M. Majmundar of AT{\&}T for his valuable feedback and discussions on the problem and the results, and R. M. Dreifuerst of NC State University for his gracious assistance in implementing their code.

\ifCLASSOPTIONcaptionsoff
  \newpage
\fi

\end{document}